\documentclass{jfm}

\usepackage{amsmath}
\usepackage{amsfonts}
\usepackage{amssymb}
\usepackage{graphicx}
\usepackage{caption}
\usepackage{subcaption}
\usepackage{mathtools}
\usepackage{natbib}
\usepackage[capitalise]{cleveref}

\newcommand{\pder}[2] {\frac{\partial #1}{\partial #2}}
\newcommand{\dder}[2] {\frac{\mathrm{d} #1}{\mathrm{d} #2}}
\newcommand{\pderline}[2] {{\partial #1}/{\partial #2}}
\newcommand{\dint}{\,\textrm{d}}
\newcommand{\mat}{\bf}
\newcommand{\uv}[1] {\hat{\textbf{#1}}}

\DeclareMathOperator{\J}{J}
\DeclareMathOperator{\R}{R}
\DeclareMathOperator{\Y}{Y}
\let\H\relax\DeclareMathOperator{\H}{H}

\title{The evolution of surface quasi-geostrophic modons on sloping topography}
\author{Matthew N. Crowe and Edward R. Johnson}
\date{}

\shorttitle{SQG modons on sloping topography}
\shortauthor{M. N. Crowe \& E. R. Johnson}
\affiliation{Department of Mathematics, University College London, London, WC1E 6BT, UK}
\date{}

\begin{document}

\maketitle

\begin{abstract}
This work discusses modons, or dipolar vortices, propagating along sloping topography. Two different regimes exist which are studied separately using the surface quasi-geostrophic equations. First, when the modon propagates in the opposite direction to topographic Rossby waves, steady solutions exist and a semi-analytical method is presented for calculating these solutions. Second, when the modon propagates in the same direction of the Rossby waves, a wave wake is generated. This wake removes energy from the modon causing it to decay slowly. Asymptotic predictions are presented for this decay and found to agree closely with numerical simulations. Over long times, decaying vortices are found to break down due to an asymmetry resulting from the generation of waves inside the vortex. A monopolar vortex moving along a wall is shown to behave in a similar way to a dipole, though the presence of the wall is found to stabilise the vortex and prevent the long-time breakdown. The problem is mathematically equivalent to a dipolar vortex moving along a density front hence our results apply directly to this case.
\end{abstract}


\section{Introduction}

Modon, or dipolar vortices, are exact analytical solutions to many geophysical problems and are often used to model a pair of co-propagating, oppositely signed vortices \citep{stern_1975,FLIERLHAINES94,MurakiS07}. Physically, these vortex solutions occur in both the ocean and atmosphere and are important for transporting fluid across large distances \citep{HughesM17,NiZWH20}. In particular, modons are able to travel in directions that Rossby waves cannot, enabling a transfer of information that is not captured by linear wave models.

Recent analysis of satellite observations and Argo float data by \citet{NiZWH20} indicates that surface modons are common features of the upper ocean, especially in energetic regions such as the Southern Ocean and the Gulf Stream. These modons were found to enhance the Ekman pumping velocity, leading to a strong vertical exchange of heat, carbon and nutrients between the ocean surface and interior. In addition to surface vortices, vortex structures may also appear on bottom boundaries. Examples of these deep-ocean vortices include lenses generated by boundary currents \citep{nof_1991} and eddy trains generated by the blocking of coastal-trapped waves \citep{rodney_johnson_2014}. These vortices may form dipolar structures through the coupling of oppositely signed vortices or through the image effect imposed by a coastal boundary \citep{shi_nof_1994}. Modons are similarly important in the atmosphere, where they may be involved in a variety of processes, such as atmospheric blocking \citep{mcwilliams_1980} and Madden-Julian Oscillation events \citep{rostami_zeitlin_2021}.

The ocean and atmosphere support waves across a wide range of scales and hence moving vortices may generate a wave wake. Some families of waves---such as Rossby waves and coastal-trapped waves---travel in one direction only, resulting in an asymmetry between vortices moving with and against the wave direction. For example, on a beta-plane, eastward and westward propagating vortices behave differently \citep{FLIERLHAINES94}. As a vortex generates waves, it  loses energy, leading to a gradual change in its the speed and structure. \citet{FLIERLHAINES94} describe this decay by wave radiation of beta-plane vortices. A general framework for this process has recently been formulated and applied to a range of dipolar vortex problems  \citep{JohnsonC21,crowe_johnson_2021}.

Here we consider the evolution of modon moving along a sloping bottom boundary in the direction perpendicular to the slope. We use a 3D, quasi-geostrophic model and make the surface QG (SGQ) assumption that the interior vorticity vanishes and hence the dynamics are controlled by the advection of density on the bottom boundary \cite[][]{Johnson78b}. This system is mathematically equivalent to the case of a background horizontal density gradient \citep{HELD_ET_AL_1995} and hence our results also apply to this problem, relevant both to motions in the atmosphere and the ocean.

\cref{sec:setup} notes the governing equations. \cref{sec:steady_sol} considers the case of a vortex moving in the opposite direction to the Rossby waves, described here as retrograde motion. Here no wave wake is formed and steady vortex solutions exist. These solutions are those found by a nonlinear numerical approach in \citet{MurakiS07} but obtained here by a semi-analytical method that allows solutions to be found by solving a standard linear eigenvalue problem. \cref{sec:decay_sol} considers modons moving in the same direction as the Rossby waves, described as prograde. Here a wave wake which removes energy from the vortex is generated. We present asymptotic predictions for the decay of the vortex in the limit of shallow slope. \cref{sec:num_sol} gives numerical tests of the predictions showing them to be valid for a range of order one slope parameters. \cref{sec:breakdown} describes the long term behaviour of the decaying vortex solutions showing that the modon structure eventually breaks down. \cref{sec:discuss} briefly discusses the results.

\section{Setup}
\label{sec:setup}

We consider a semi-infinite layer of rotating, stratified fluid over a slope with constant gradient. We assume that the slope is sufficiently small for the motion to be governed by the three-dimensional quasi-geostrophic (QG) equations and introduce  coordinates moving at speed $U$ in the direction ($x$) perpendicular to the down-slope direction ($y$). Finally, we make the surface QG (or SQG) approximation that the interior potential vorticity in the layer vanishes. Our system then satisfies
\begin{equation}
\label{eq:lap}
\nabla^2 \psi = 0\quad\textrm{for}\quad z>0,
\end{equation}
subject to bottom boundary condition
\begin{equation}
\label{eq:SQG_BC}
\left(\pder{}{t}-U\pder{}{x}+J\left[\psi,*\right]\right)\pder{\psi}{z}+\lambda\pder{\psi}{x} = 0\quad\textrm{on}\quad z = 0,
\end{equation}
and far-field condition that $\psi\to 0$ as $z \to \infty$. Here $J[f,g] = \partial_x f\, \partial_y g - \partial_x g\, \partial_y f$ is the Jacobian derivative and $\lambda$ represents the slope gradient in the QG limit. This system is discussed in \cite{rodney_johnson_2014} and is mathematically equivalent to the case of an SQG layer in a background buoyancy gradient, with $\lambda$ representing the constant horizontal buoyancy gradient \citep{HELD_ET_AL_1995}. The results below thus apply to modons moving along a density front in the atmosphere or ocean.

\section{Steady retrograde modons, $\mu=\lambda a/U\geq0$.}
\label{sec:steady_sol}

We now seek steady solutions in the form of dipolar vortices (or modons) travelling at speed $U$ in the cross-slope ($x$) direction. We consider vortices which are circular in the plane $z=0$ with (cylindrical) radius $r = a$. Taking $\partial_t = 0$, \cref{eq:SQG_BC} may be written as
\begin{equation}
J\left[\psi+Uy,\pder{\psi}{z}+\lambda y\right] = 0\quad \textrm{on}\quad z = 0,
\end{equation}
which gives that
\begin{equation}
\pder{\psi}{z}+\lambda y = F\left(\psi+Uy\right)\quad \textrm{on}\quad z = 0,
\end{equation}
for some arbitrary function, $F$. We will consider the class of solutions where $F$ is a linear function to obtain the Long's model system
\begin{equation}
\label{eq:gov_eq1}
\pder{\psi}{z}+\lambda y = -\frac{K}{a}\left(\psi+Uy\right)\quad \textrm{on}\quad z = 0,
\end{equation}
where the value of $K$ may be different inside and outside the vortex. Far from the vortex, we require that both $\psi$ and $\partial_z \psi$ vanish and hence $K/a = -\lambda/U$ outside the vortex ($r > a$). Conversely, inside the vortex ($r < a$) $K$ must be positive to ensure that $\psi$ decays away from the boundary but will otherwise remain arbitrary until determined during the solution. Continuity requires that $\psi+Uy = 0$ on the vortex boundary, $r = a$. Therefore, \cref{eq:gov_eq1} becomes
\begin{equation}
\label{eq:gov_eq2}
\pder{\psi}{z} =\begin{cases}
-\frac{K}{a}\psi-\left(\frac{UK}{a}+\lambda\right) y & \textrm{for}\quad z = 0,\, r < a, \\
\;\;\;\frac{\lambda}{U}\psi & \textrm{for}\quad z = 0,\, r > a.
\end{cases}
\end{equation}

Steady modon solutions are now obtained by seeking solutions that satisfy \cref{eq:lap,eq:gov_eq2}. Following \cite{MurakiS07} and \citet{JohnsonC22} we write
\begin{equation}
\label{eq:sol_psi}
\psi(x,y,z) = Ua \sin\theta \int_0^\infty \hat{\psi}(\xi) \J_1(\xi r/a)\exp(-\xi z/a) \,\xi \dint \xi,
\end{equation}
where $\theta$ is the (cylindrical) polar angle, $\J_1$ denotes the Bessel function of the first kind of order $1$ and $\hat{\psi}$ describes the Hankel transform of $\psi$ for $\theta=\pi/2$ and $z = 0$. Substituting \cref{eq:sol_psi} into \cref{eq:gov_eq2} gives
\begin{subequations}
\begin{align}
\int_0^\infty \hat{\psi}(\xi) \J_1(\xi s) \,\xi^2 \dint \xi - K\int_0^\infty \hat{\psi}(\xi) \J_1(\xi s) \,\xi \dint \xi=&  \left(K+\mu\right) s & \quad\textrm{for}\quad s < 1,
\label{eq:int1}
\\
\int_0^\infty \hat{\psi}(\xi) \J_1(\xi s) \,\xi^2 \dint \xi + \mu\int_0^\infty \hat{\psi}(\xi) \J_1(\xi s) \,\xi \dint \xi=&  \;0 & \quad\textrm{for}\quad s > 1,
\label{eq:int2}
\end{align}
\end{subequations}
where $s = r/a$ is the rescaled vortex radius and $\mu = \lambda a/U$ is the rescaled slope. We now substitute
\begin{equation}
A(\xi) = \left(\xi^2+\mu\xi\right)\hat{\psi}(\xi),
\end{equation}
so \cref{eq:int2} becomes
\begin{equation}
\int_0^\infty A(\xi) \J_1(\xi s) \dint \xi = 0\quad\textrm{for}\quad s > 1.
\end{equation}
At this point it is convenient to expand $A(\xi)$ in terms of Bessel functions as
\begin{equation}
A(\xi) = \sum_{n = 0}^\infty a_n \J_{2n+2}(\xi),
\end{equation}
for some undetermined coefficients, $a_n$ \citep{Tranter71,JohnsonC22}. This form allows us to exploit the integral relationship
\begin{equation}
\label{eq:Rn_def}
\int_0^\infty \J_{2n+2}(\xi)\J_1(\xi s) \dint\xi = \begin{cases}
\R_n(s) &\textrm{for}\quad s<1, \\
0 &\textrm{for}\quad s>1,
\end{cases}
\end{equation}
such that \cref{eq:int2} is automatically satisfied for all choices of $a_n$. Here $\R_n(s)$ denotes the Zernike Radial function \citep{BornW19}, a set of degree $2n+1$ polynomials orthogonal over $s\in[0,1] $ with weight $s$. Substituting for $A$ in \cref{eq:int1} gives
\begin{equation}
\int_0^\infty \frac{\xi-K}{\xi+\mu}A(\xi) \J_1(\xi s) \dint \xi = (K+\mu) s \quad\textrm{for}\quad s < 1,
\end{equation}
hence
\begin{equation}
\label{eq:int3}
\sum_{n=0}^\infty a_n \int_0^\infty \frac{\xi-K}{\xi+\mu}\J_{2n+2}(\xi) \J_1(\xi s) \dint \xi = (K+\mu) s \quad\textrm{for}\quad s < 1.
\end{equation}
We now multiply \cref{eq:int3} through by $s \R_m(s)$ and integrate over $s \in [0,1]$. Note that, from \cref{eq:Rn_def}, $\R_n(s)$ is the Hankel transform of $\J_{2n+2}(\xi)/\xi$ so inverting this relation gives
\begin{equation}
\label{eq:J_rel}
J_{2m+2}(\xi)/\xi = \int_0^1 s \R_m(s) \J_1(\xi s) \dint s.
\end{equation}
Using \cref{eq:J_rel}, we obtain
\begin{equation}
\label{eq:int4}
\sum_{n = 0}^\infty a_n \int_0^\infty \frac{\xi-K}{\xi(\xi+\mu)} \J_{2n+2}(\xi)\J_{2m+2}(\xi) \dint \xi = \frac{K+\mu}{4} \delta_{m 0},
\end{equation}
where $m \in \{0,1,2,\dots\}$ and $\delta_{ij}$ is the Kronecker delta. \cref{eq:int4} takes the form of an infinite, inhomogeneous, generalised eigenvalue problem of the form
\begin{equation}
\label{eq:eig1}
\left({\mat A}-K{\mat B}\right)\textbf{a} = \textbf{c}.
\end{equation}
Here $\textbf{a}$ denotes the $a_n$,
\begin{equation}
\label{eq:A_def}
[{\mat A}]_{mn} = A_{mn} = \int_0^\infty \frac{1}{\xi+\mu}\J_{2m+2}(\xi)\J_{2n+2}(\xi) \dint \xi,
\end{equation}
\begin{equation}
\label{eq:B_def}
[{\mat B}]_{mn} = B_{mn} = \int_0^\infty
\frac{1}{\xi(\xi+\mu)} \J_{2m+2}(\xi) \J_{2n+2}(\xi) \dint \xi,
\end{equation}
and
\begin{equation}
[\textbf{c}]_m = c_m = \frac{K+\mu}{4}\delta_{m0}.
\end{equation}
The integrals in \cref{eq:A_def,eq:B_def} are oscillatory and hence difficult to calculate directly. A method of re-writing these integrals in a non-oscillatory form that can be easily calculated is discussed in \cref{sec:bessel_eval}.

The inhomogeneity of \cref{eq:eig1} is confined to the row $m = 0$ and may be removed by considering the boundary conditions on $s = 1$. By \cref{eq:gov_eq2} we require
\begin{equation}
\pder{\psi}{z} = \frac{\lambda}{U}\psi, \quad\textrm{on}\quad s = 1,
\end{equation}
therefore
\begin{equation}
\sum_{n = 0}^\infty a_n \int_0^\infty \J_{2n+2}(\xi)\J_1(\xi)\dint\xi = 0,
\end{equation}
and hence
\begin{equation}
\label{eq:a0_def}
a_0 = \sum_{n=1}^\infty (-1)^{n+1}a_n.
\end{equation}
We may therefore remove the first row of \cref{eq:eig1} and replace all instances of $a_0$ using \cref{eq:a0_def}. The resulting linear system allows us to determine the $a_n$ up to a multiplicative constant which may be subsequently determined using the first row of \cref{eq:eig1}. In the case of $\mu = 0$, the integrals $A_{mn}$ and $B_{mn}$ may be found analytically as shown in \citet{JohnsonC22}. Our resulting linear system is
\begin{equation}
\left[ {\mat B} - \frac{1}{K}\, {\mat A} \right] \textbf{a} = \boldsymbol 0,
\end{equation}
where
\begin{equation}
\tilde{A}_{mn} = A_{mn}+(-1)^{n+1} A_{m0},
\end{equation}
and
\begin{equation}
\tilde{B}_{mn} = B_{mn}+(-1)^{n+1} B_{m0},
\end{equation}
for $m,n \in \{1,2,\dots\}$ and the eigenvectors, $a_n$, are scaled using the condition
\begin{equation}
\left(A_{00}-K B_{00}\right)a_0 + \sum_{n=1}^\infty\left(A_{0n}-K B_{0n}\right)a_n = \frac{K+\mu}{4},
\end{equation}
where $a_0$ is given by \cref{eq:a0_def}. This linear system may be truncated to a finite number of terms, $m,n \in \{1,2,\dots,N\}$, to obtain numerical results using any eigensolver. We solve for $N$ eigenvalues, $K$, and sets of coefficients, $a_n$, with the smallest value of $K$ corresponding to the required dipolar vortex. Solutions with higher $K$ are valid and correspond to vortex solutions with a higher order radial structure. \cite{JohnsonC22} show that for $\lambda=0$ the higher order solutions are unstable through pairing of the inner vortices. Higher order  solutions for $\lambda>0$ can be expected to be susceptible to the same instability so will not be considered further. This method allows modon solutions to be found quickly and accurately and a value of $N = 19$ is sufficient to ensure that the largest neglected coefficient is less than $10^{-6}$. Our solution satisfies
\begin{equation}
\label{eq:surf_bc}
\left[\pder{\psi}{z}-\frac{\lambda}{U}\psi\right]_{z = 0} = \begin{cases}
-U \sin\theta \sum_{n=0}^\infty a_n \R_n(r/a) & \textrm{for}\quad r<a,
\\
0 & \textrm{for}\quad r>a,
\end{cases}
\end{equation}
therefore, numerically, once the $a_n$ are determined, it is often easier to solve $\nabla^2 \psi = 0$ subject to \cref{eq:surf_bc} than evaluate the Hankel transforms in \cref{eq:sol_psi}. Note also that the Zernike sum in \cref{eq:surf_bc} can be evaluated from a three-term recurrence relation without requiring the form of the Zernike polynomials \citep{JohnsonC22}. The solutions for $\lambda\geq0$ are the same as the solutions of \citet{MurakiS07}, however the approach here allows solutions to be found using a compact semi-analytical approach rather than the non-linear root-finding method of \citet{MurakiS07}.

\begin{figure}
	\centering
	\begin{subfigure}[b]{\textwidth}
	\centering
	\includegraphics[trim={0cm 0cm 0cm 0cm},clip,width=0.49\textwidth]{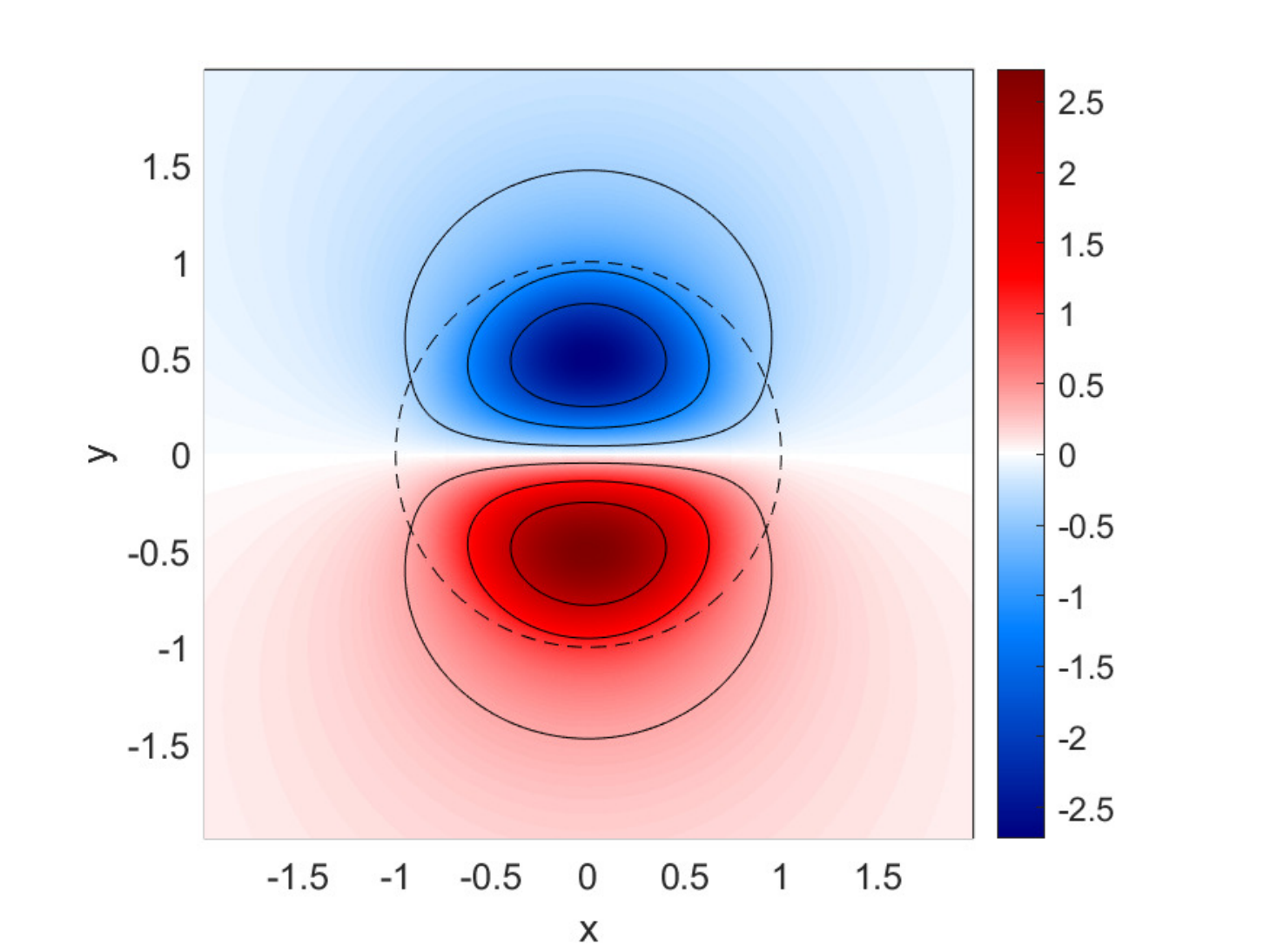}\hfill\includegraphics[trim={0cm 0cm 0cm 0cm},clip,width=0.49\textwidth]{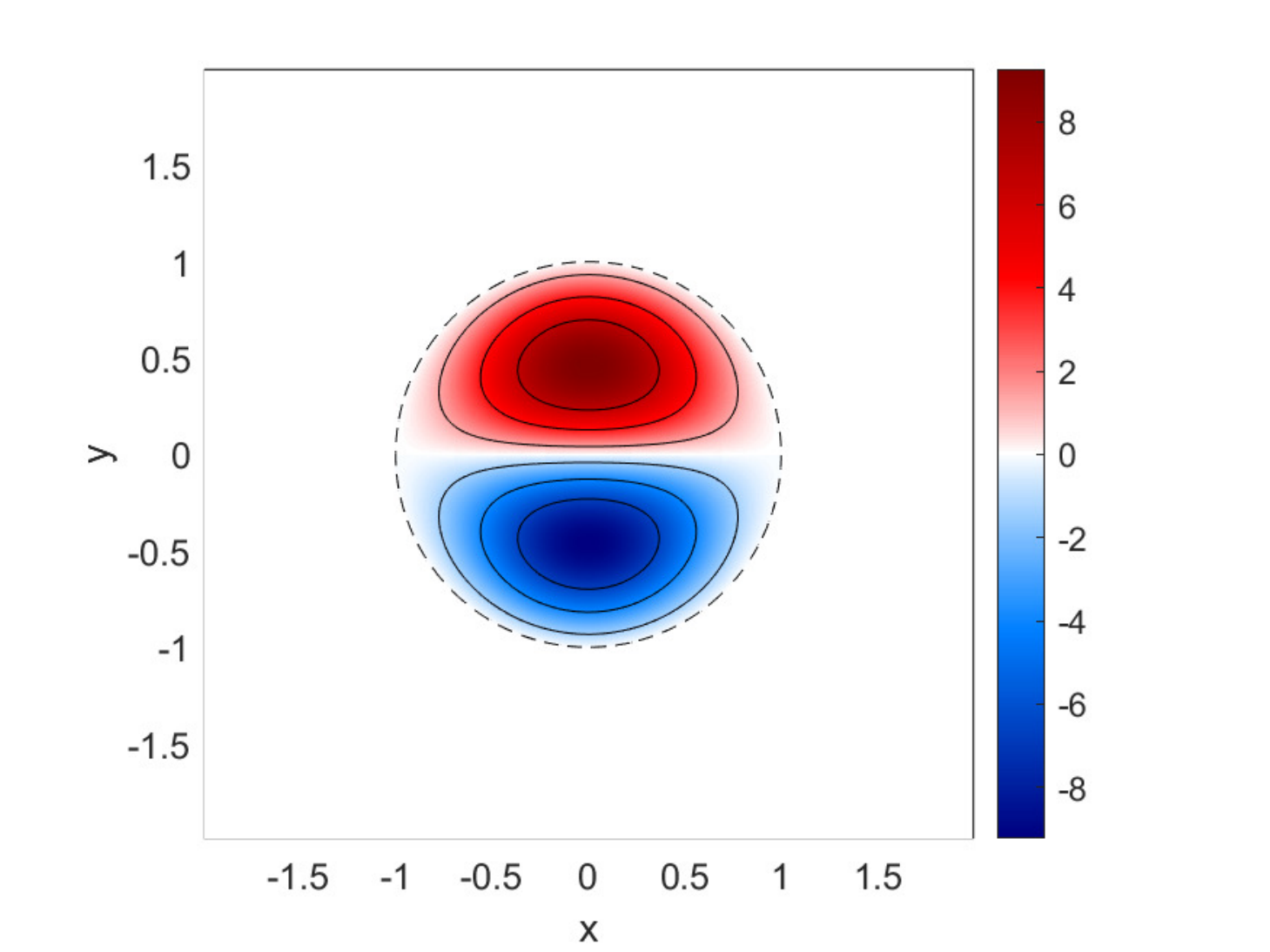}
	\caption{}
	\end{subfigure}
	\begin{subfigure}[b]{\textwidth}
	\centering
	\includegraphics[trim={0cm 0cm 0cm 0cm},clip,width=0.49\textwidth]{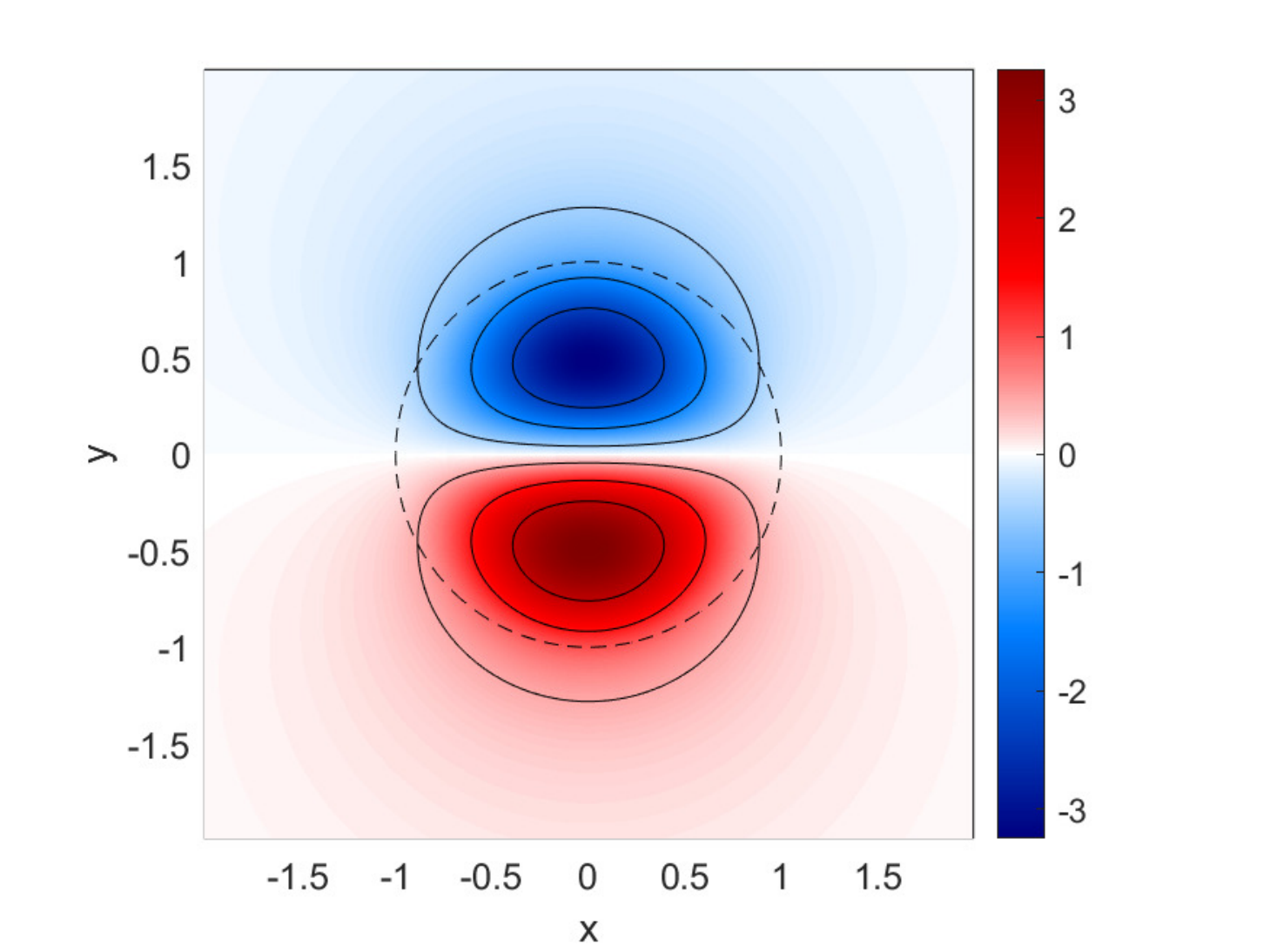}\hfill\includegraphics[trim={0cm 0cm 0cm 0cm},clip,width=0.49\textwidth]{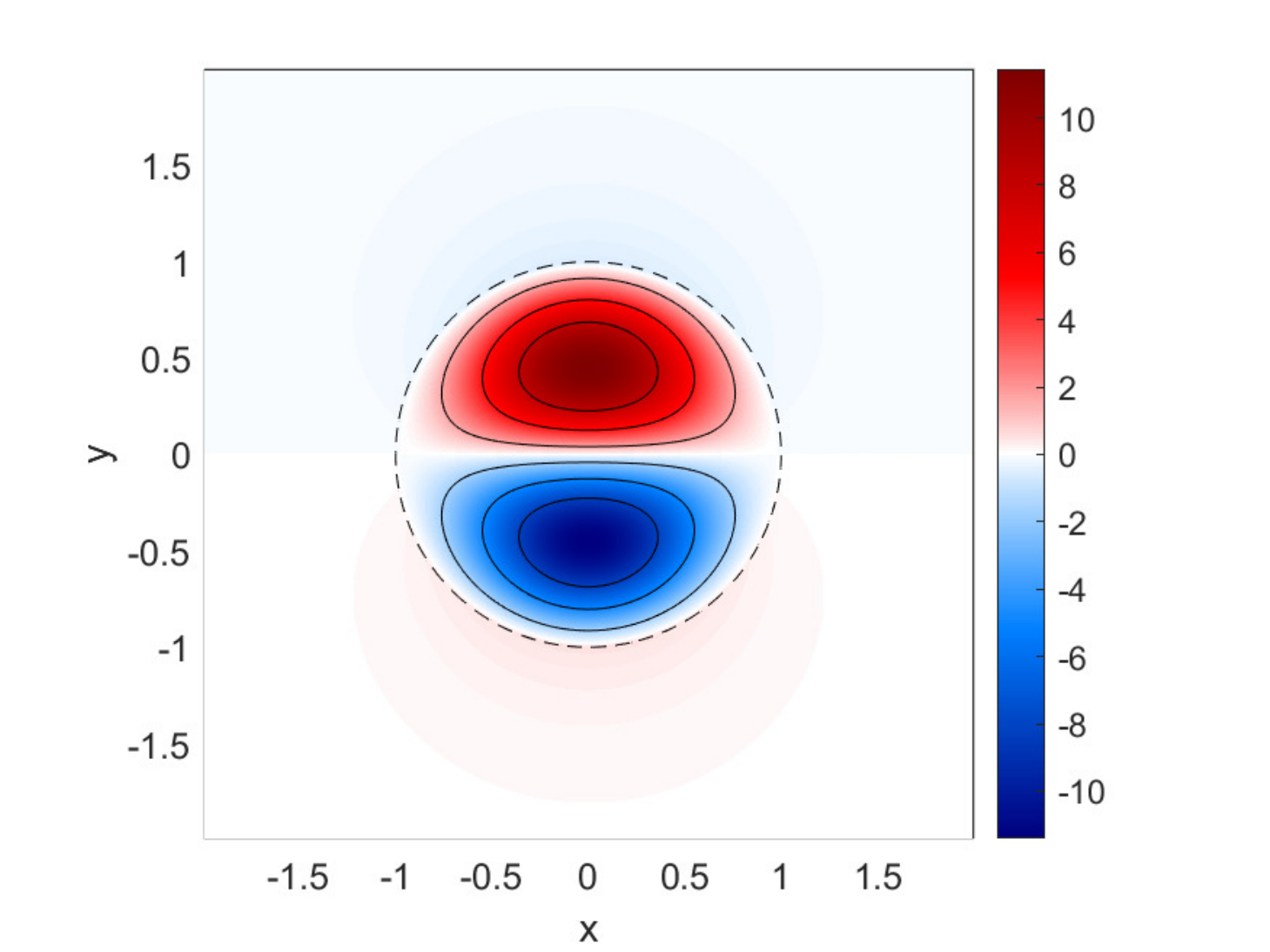}
	\caption{}
	\end{subfigure}
	\begin{subfigure}[b]{\textwidth}
	\centering
	\includegraphics[trim={0cm 0cm 0cm 0cm},clip,width=0.49\textwidth]{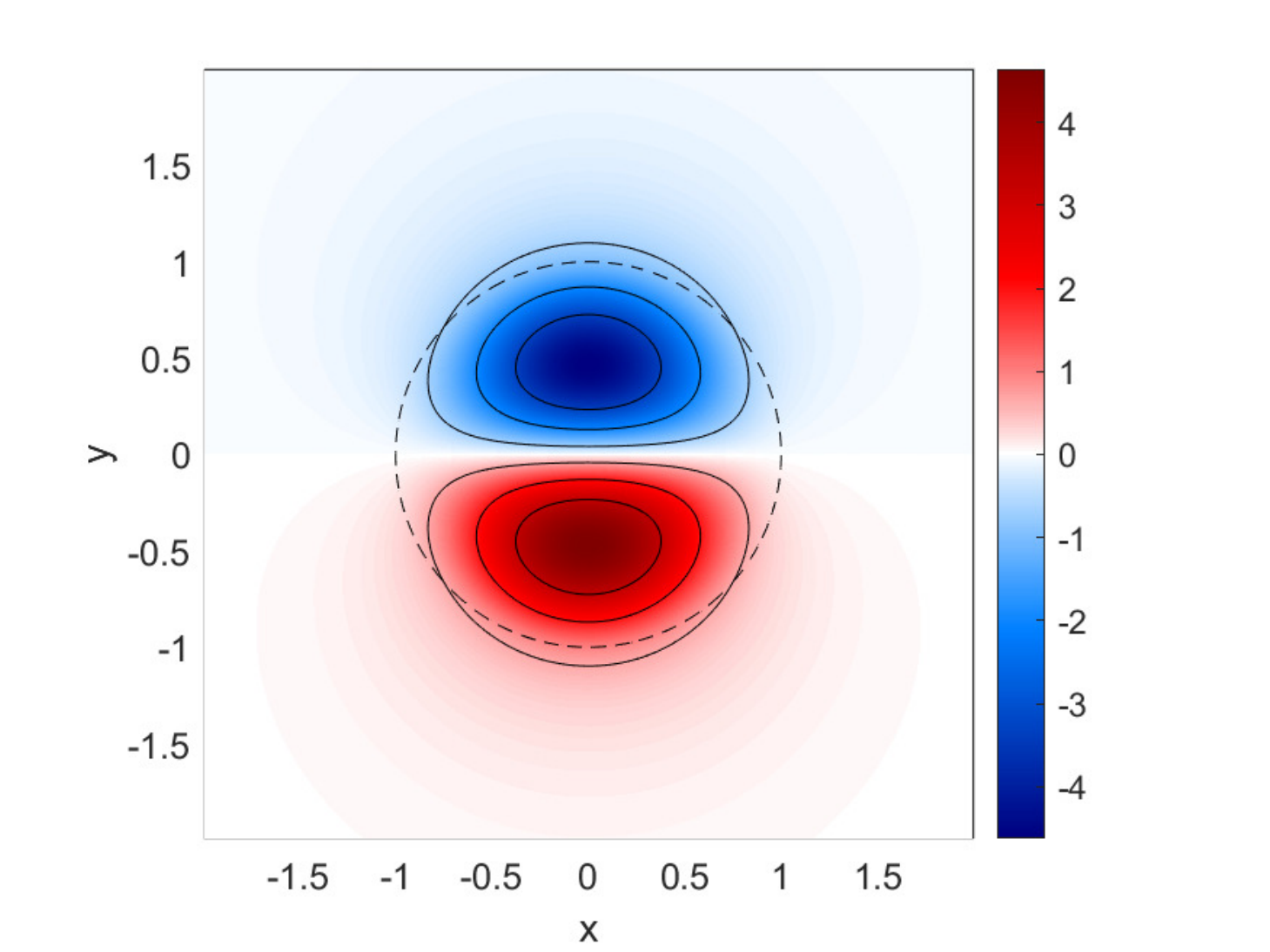}\hfill\includegraphics[trim={0cm 0cm 0cm 0cm},clip,width=0.49\textwidth]{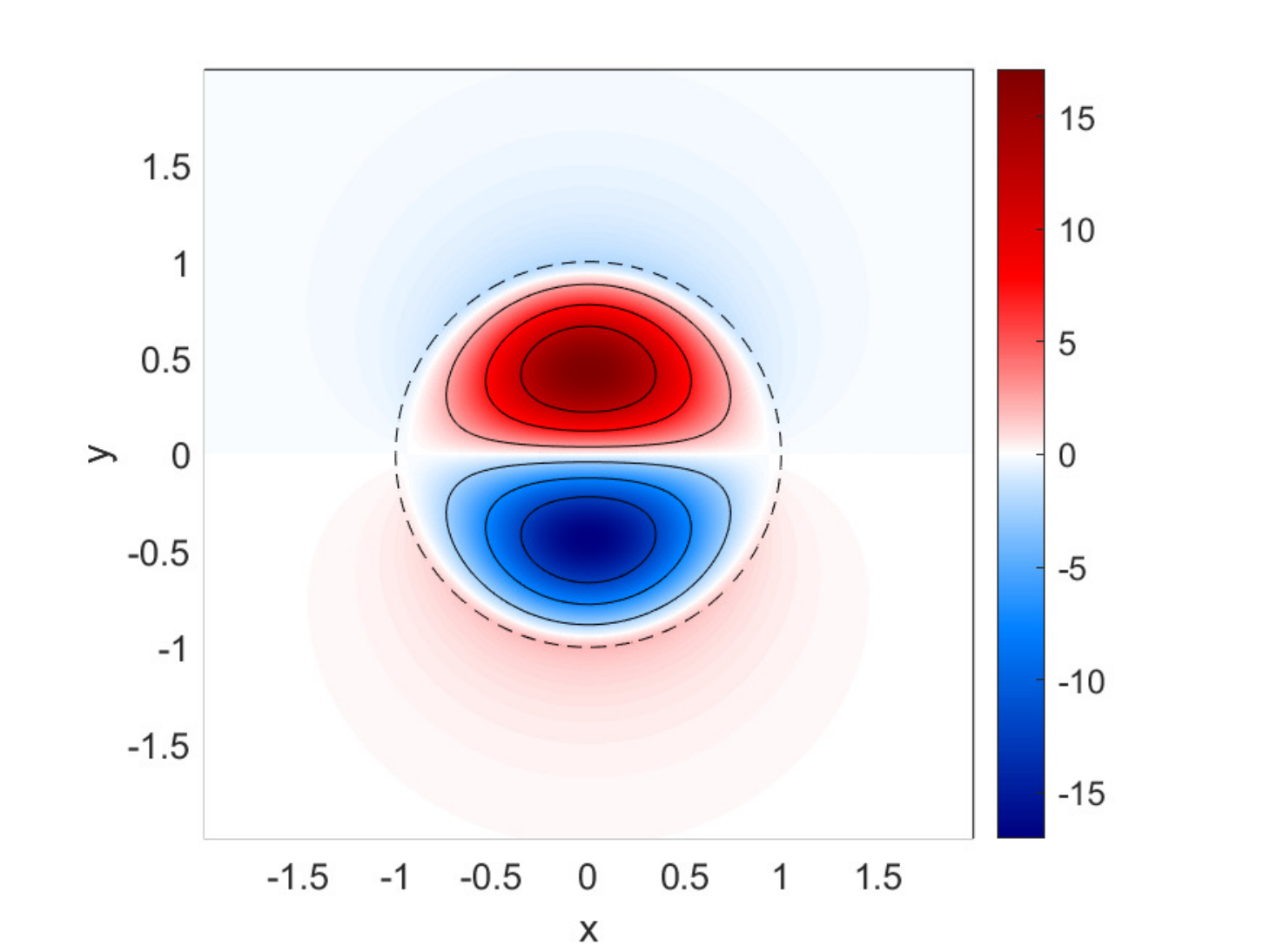}
	\caption{}
	\end{subfigure}
	\caption{Plots for $\psi$ (left) and $\psi_z$ (right) on $z = 0$ for a range of values of $\lambda$; (a) $\lambda = 0$, (b) $\lambda = 0.5$ and (c) $\lambda = 2$. We use $(U,a) = (1,1)$ throughout. The dashed line denotes the vortex boundary, $r = a$.}
    \label{fig:vort_sol}
\end{figure}

\cref{fig:vort_sol} shows $\psi$ and $\pderline{\psi}{z}$ for a range of non-negative values of $\lambda$. We observe that the maximum values of both $\psi$ and $\pderline{\psi}{z}$ increase with increasing $\lambda$. Further, in the plots of $\pderline{\psi}{z}$ we observe regions just outside $r = a$ of oppositely signed values to those within $r < a$. This is due to the relation $\pderline{\psi}{z} = \lambda \psi /U$ for $r > a$ and corresponds to the vortex being surrounded by regions of fluid with an oppositely signed buoyancy to the peaks. Evolving these solutions forward in time using the method discussed in \cref{sec:num_sol} shows that they are both steady and stable over long times.

\begin{figure}
	\centering
	\begin{subfigure}[b]{0.33\textwidth}
	\centering
	\includegraphics[trim={0cm 0cm 0cm 0cm},clip,width=\textwidth]{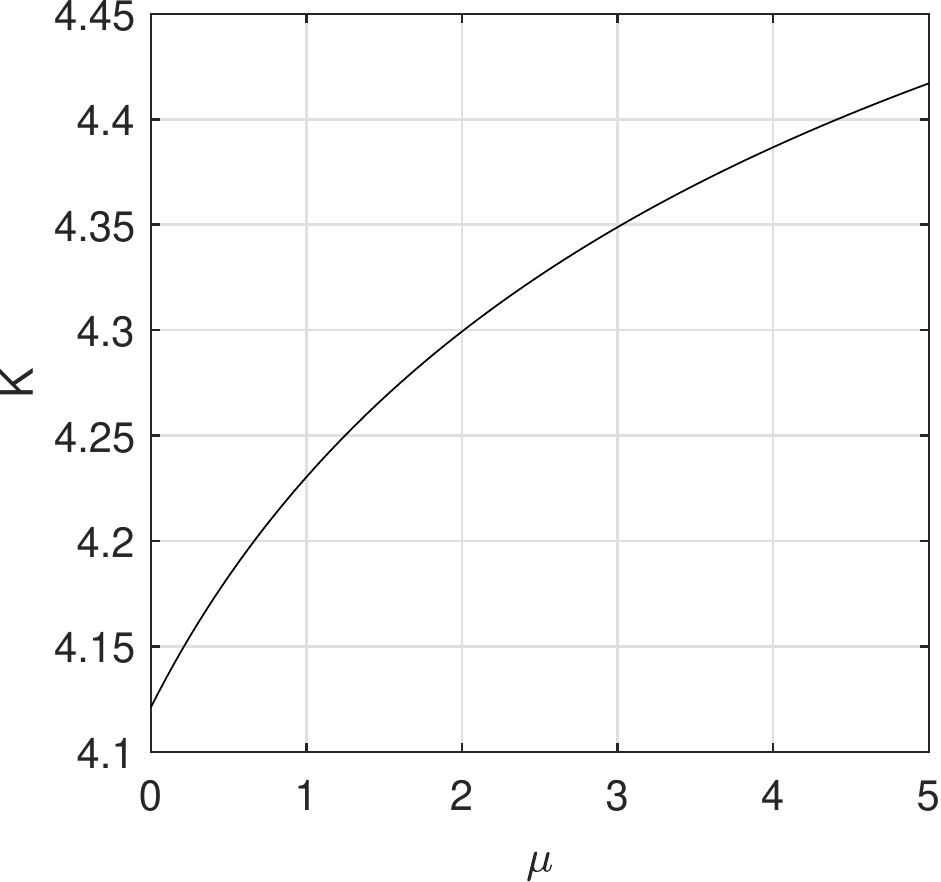}
	\caption{}
	\end{subfigure}
	\begin{subfigure}[b]{0.32\textwidth}
	\centering
	\includegraphics[trim={0cm 0cm 0cm 0cm},clip,width=\textwidth]{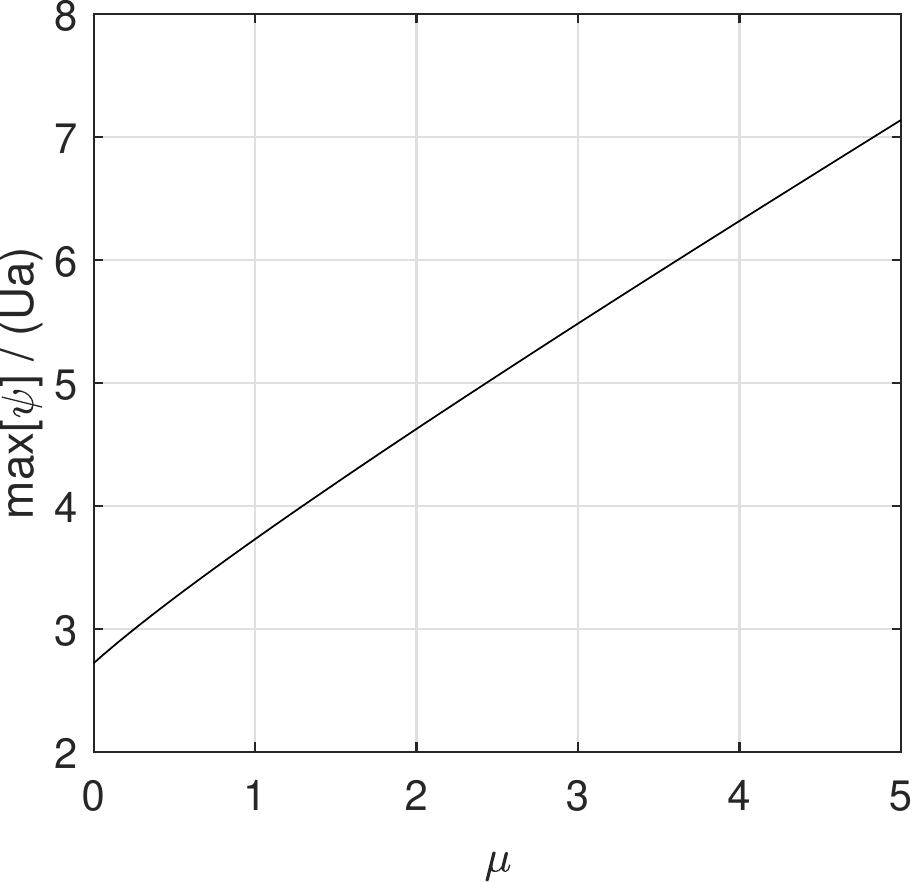}
	\caption{}
	\end{subfigure}
	\begin{subfigure}[b]{0.33\textwidth}
	\centering
	\includegraphics[trim={0cm 0cm 0cm 0cm},clip,width=\textwidth]{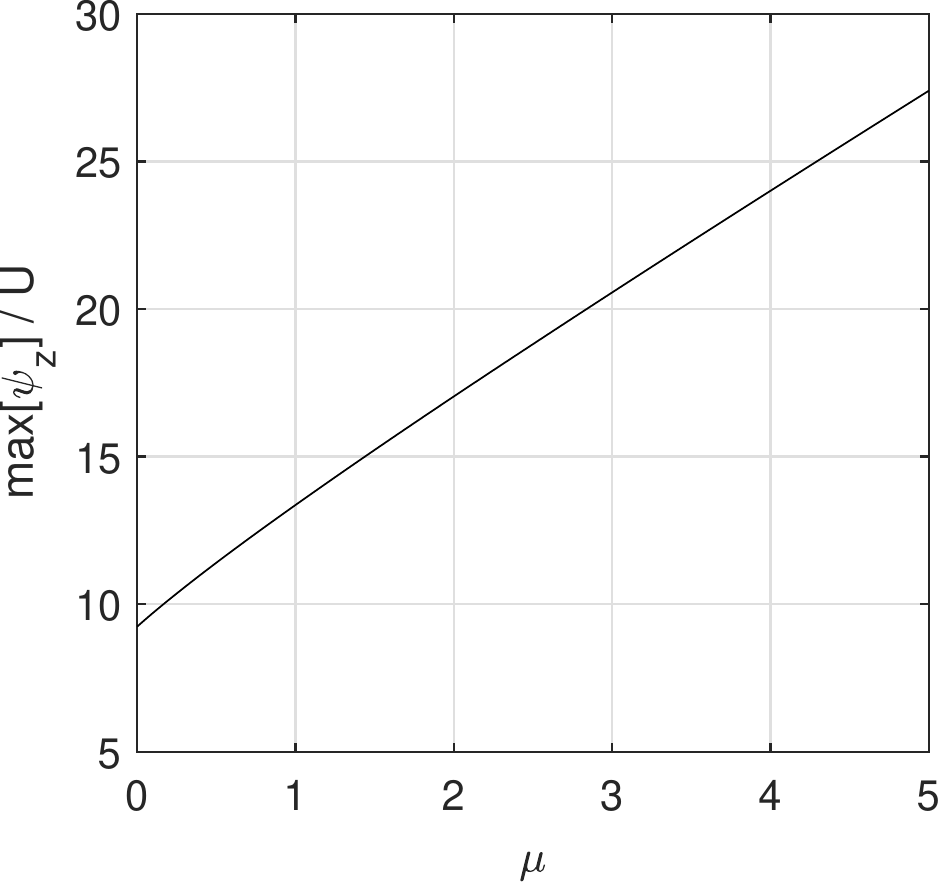}
	\caption{}
	\end{subfigure}
	\caption{(a) $K$, (b) $\max[\psi]_{z = 0}$, and (c) $\max[\pderline{\psi}{z}]_{z = 0}$ as functions of $\mu = \lambda a/U$.}
    \label{fig:mu_dep}
\end{figure}

We note that the maximum values of $\psi/(Ua)$ and $[\pderline{\psi}{z}]/U$ on $z = 0$ depend only on $U$ and $a$ through the parameter $\mu$. This parameter dependence occurs as the integral in \cref{eq:sol_psi} depends only on $r/a$, $z/a$ and $\mu$ as $\hat{\psi}$ depends only on $\mu$. \cref{fig:mu_dep} shows $K$, $\max[\psi]_{z = 0}$, and $\max[\pderline{\psi}{z}]_{z = 0}$ as functions of non-negative $\mu$. All quantities are observed to increase with $\mu$.

\section{Decaying prograde modons, $\mu=\lambda a/U<0$.}
\label{sec:decay_sol}

If a vortex has $\mu < 0$ then the integrals in \cref{eq:A_def,eq:B_def} are singular at $\xi = |\mu|$ and do not converge. This corresponds to the case where the moving vortex generates a topographic Rossby wave wake. Here, solutions to the governing equations are not unique unless a radiation condition is applied \citep{Croweetal20,crowe_johnson_2021} to ensure that no wave energy is entering the flow from the far field.

To study the case of $\mu < 0$, we follow the approach of \citet{FLIERLHAINES94} and \citet{JohnsonC21}. We begin by determining the wave field generated by a moving dipolar vortex. The energy flux into this wave field is determined and then equated to the loss of vortex energy to obtain a prediction for the vortex evolution in the case of small $\mu$.

\subsection{The wave-field solution}

In the far field, the vortex resembles a point dipole at the origin so the wavefield satisfies
\begin{equation}
\label{eq:wave_1}
\pder{\psi}{z}-\frac{\lambda}{U}\psi = -m\, 
\delta(x)\delta'(y)  \quad\textrm{on}\quad z = 0,
\end{equation}
and
\begin{equation}
\label{eq:wave_2}
\nabla^2 \psi = 0 \quad\textrm{in}\quad z > 0,
\end{equation}
where $m$ denotes the dipole impulse. We now write $\kappa = -\lambda/U$ or equivalently $\kappa = -\mu/a$, so $\kappa$ will correspond to the wavenumber of the generated wave-field. \cref{eq:wave_1,eq:wave_2} can be solved using a two-dimensional Fourier transform in $x$ and $y$ to obtain
\begin{equation}
\label{eq:psi_wave_sol}
\psi = \frac{m}{4\pi^2}\iint \frac{il}{\sqrt{k^2+l^2}-\kappa} \exp\left[ ik x + ily - z\sqrt{k^2+l^2}\right] \dint k \dint l.
\end{equation}
For $\kappa>0$ (and hence $\mu < 0$), there is a singularity in the integrand of \cref{eq:psi_wave_sol} at $k^2+l^2 = \kappa^2$. This singularity corresponds to the appearance of topographic Rossby waves and the integral must be evaluated such that the solution obeys the radiation condition of no disturbances propagating inwards from the far-field. Our dipole solution may be written as the $y$ derivative of the monopole solution of \cite{Johnson78c} so
\begin{equation}
\psi = m\pder{G}{y},
\end{equation}
where
\begin{equation}
G(r,\theta,z) \!=\! \frac{1}{2\pi\sqrt{r^2\!+\!z^2}} \!-\! \frac{\kappa}{4}\!\exp(-\kappa z)\!\!\left[ \mathcal{H}_0(\kappa r) \!+\! \Y_0(\kappa r) \!+\! \frac{2}{\pi}\!\!\int_0^{\kappa z}\!\!\!\! \frac{\exp(t)}{\sqrt{t^2+\kappa^2 r^2}} \dint t +\! \frac{8}{\pi} S(\kappa r,\theta)\!\right]\!,\!\!
\end{equation}
$\mathcal{H}_0$ is the zero-order Struve function and $\Y_0$ is a Bessel function of the second kind. The final term, $S$, describes the wavefield and is given by
\begin{equation}
S(\kappa r,\theta) = \sum_{k=0}^\infty \frac{\cos[(2k+1)\theta] J_{2k+1}(\kappa r)}{2k+1}.
\end{equation}

\subsection{The wave-energy flux}

We now reduce to the case of small $\kappa$ and consider a semi-infinite cylinder of radius $R\gg 1$ centred on the vortex.As $\kappa$ is small, we take $R$ such that $\kappa R \ll 1$ and note that the outward energy flux through this cylinder must be independent of $R$. We are therefore able to calculate this energy flux using the streamfunction in the limit of small $\kappa r$. Using polar coordinates
\begin{equation}
\psi = m\left[\sin\theta\pder{G}{r}+\frac{\cos\theta}{r}\pder{G}{\theta}\right],
\end{equation}
and hence
\begin{equation}
\label{eq:psi_asymp}
\psi \sim \frac{m}{\pi}\!\left(\! \sin\theta\!\left[-\frac{r}{2(r^2+z^2)^{3/2}} + O(\kappa)\right]\! + \sin\theta\cos\theta\!\left[ \frac{\kappa^4}{3} r^2 \exp(-\kappa z) \!+ O(\kappa^6)\right]\! + O(\kappa^6)  \!\right)\!.
\end{equation}
The rate of change of vortex energy is given by the sum of the work done by the pressure on the cylinder and the energy advected through the cylinder. This energy loss is calculated in \cref{sec:E_evol} and given by \cref{eq:E_flux}. Using our asymptotic result from \cref{eq:psi_asymp} we can calculate the rate of energy loss in the limit of small $\mu$ as
\begin{equation}
\dder{E}{t} = -\frac{m^2 |U| \kappa^4}{3\pi} = -\frac{\pi |a_0|^2 \lambda^4 a^6}{48 |U|},
\end{equation}
hence
\begin{equation}
\label{eq:E_evol_res}
\dder{}{t}\left[ U^2 a^3\right] = -\frac{ |a_0| \lambda^4 a^6}{24 |U|},
\end{equation}
where $a_0$ is defined in \cref{eq:a0_def} and the results for the vortex energy, $E$, and momentum, $m$, are given in the limit of small $\mu$ in \cref{eq:m_def,eq:E_def} as
\begin{equation}
m = \frac{\pi |a_0| U a^3}{4}, \quad E = \frac{\pi |a_0| U^2 a^3}{2}.
\end{equation}

\subsection{Asymptotic predictions for the vortex decay}
\label{sec:decay_pred}

To predict the vortex decay we will assume that the vortex form remains self-similar throughout the evolution and hence depends only on its speed, $U$, and radius, $a$. Both $U$ and $a$ may vary with time and hence we need two equations involving these parameters in order to determine the evolution. \cref{eq:E_evol_res} corresponds to conservation of energy and is one such equation. Conservation of total momentum or potential vorticity (PV) might be expected to hold and give further equations linking $U$ and $a$. However, our previous work \citep{JohnsonC21} has shown that these quantities are not well conserved throughout the evolution as some fluid escapes the vortex as its radius decreases, carrying significant amounts of momentum and PV with it. This issues was resolved by following \cite{FLIERLHAINES94} and instead considering the value of a tracer within the innermost streamline. Here, the natural choice is buoyancy and we therefore assume that the buoyancy at the maximum of $\psi$ within the vortex is conserved. This gives
\begin{equation}
\dder{}{t}\left. \pder{\psi}{z}\right|_{\psi = \psi_{max}} \!\!\!\!\! = 0,
\end{equation}
and is shown in \cref{sec:num_sol} to be a reasonable assumption. Noting that in the limit of small $\mu$, $\pderline{\psi}{z} \propto U$ and is independent of $a$, we have
\begin{equation}
\dder{U}{t} = 0,
\end{equation}
throughout the vortex evolution. Combining this result with \cref{eq:E_evol_res} gives our predictions for the evolution of the vortex speed and radius as
\begin{equation}
\label{eq:decay_pred}
U(t) = U(t_0), \quad a(t) = a(t_0)\left[ 1+\frac{|a_0| a(t_0)^3 \lambda^4}{72 |U|^3}(t-t_0) \right]^{-1/3},
\end{equation}
for some initial time $t = t_0$. 

\section{Numerical solutions for decaying prograde modons}
\label{sec:num_sol}

We now test the predictions of \cref{sec:decay_sol} using numerical simulations of \cref{eq:lap,eq:SQG_BC}. Solving this system directly requires an inversion of the Laplacian at each timestep in order to calculate $\pderline{\psi}{z}$. The value of $\pderline{\psi}{z}$ on the boundary is then stepped forward using \cref{eq:SQG_BC}. This approach requires a three-dimensional computational domain and is therefore computationally expensive at high resolutions. We instead take the alternative approach of solving the 2D boundary equation and determining $\pderline{\psi}{z}$ from $\psi$ using a Dirichlet-to-Neumann operator. Working in Fourier space in $x$ and $y$ we have
\begin{equation}
\pder{^2\hat\psi}{z^2} - (k^2+l^2)\, \hat\psi = 0,
\end{equation}
hence
\begin{equation}
\left.\pder{\hat\psi}{z}\right|_{z = 0} = -\sqrt{k^2+l^2}\left.\hat\psi\right|_{z = 0},
\end{equation}
where the choice of sign comes from the requirement that $\psi\to 0$ as $z\to \infty$. We now define $\mathcal{D}$ to be a linear operator which acts in Fourier space as $\hat{\mathcal{D}} = -\sqrt{k^2+l^2}$. Therefore, $\mathcal{D}$ may be thought of as both the fractional Laplacian operator, $\mathcal{D} = -\sqrt{-\nabla^2}$, and the required Dirichlet-to-Neumann map, $\pderline{\psi}{z} = \mathcal{D}\,\psi$. We therefore solve the 2D system
\begin{equation}
\label{eq:num_eqn}
\left(\pder{}{t}-C\pder{}{x}+J\left[\psi,*\right]\right)\mathcal{D}\psi+\lambda\pder{\psi}{x} = -\nu \nabla^4 \mathcal{D}\psi \quad\textrm{on}\quad z = 0,
\end{equation}
where hyperdiffusion is included for numerical stability. Here $C$ denotes the speed of the moving coordinates hence taking $C = U$ results in the vortex remaining centred on the origin.

\cref{eq:num_eqn} is solved using spectral methods via the Dedalus package \citep{BurnsVOLB20}. We use a doubly periodic two-dimensional domain of size of $102.4 \times 102.4$ with $4096$ gridpoints in each direction and expand $\psi$ in a Fourier basis in both directions. Timestepping is performed using a $3\!-\!\epsilon$ order implicit-explicit Runge-Kutta scheme. Our simulations are initialised by placing an SQG modon at the centre of the domain using the solutions from \cref{sec:steady_sol} with $\lambda = 0$ and $(U,a) = (-1,1)$. Simulations are run for $\lambda \in \{0,0.2,0.4,0.6,0.8,1.0\}$ over the time interval $t \in [0,50]$. This time interval is chosen as it ensures that the generated waves do not have sufficient time to loop around the domain and interact with the vortex.

Initially, over a short time interval $t \in [0,t_0]$, the vortex solution adjusts to the non-zero slope parameter, $\lambda$, and begins to generate a wave field. To compare our simulations with the theoretical predictions of \cref{eq:decay_pred} we take a value of $t_0 = 5$ and calculate $a(t_0)$ and $U(t_0)$ from our numerical data. This value of $t_0$ is found to be sufficient such that the transient motions arising from the initial adjustment have decayed.

\begin{figure}
    \centering
	\begin{subfigure}[b]{0.49\textwidth}
	\centering
	\includegraphics[trim={0cm 0cm 0cm 0cm},clip,width=\textwidth]{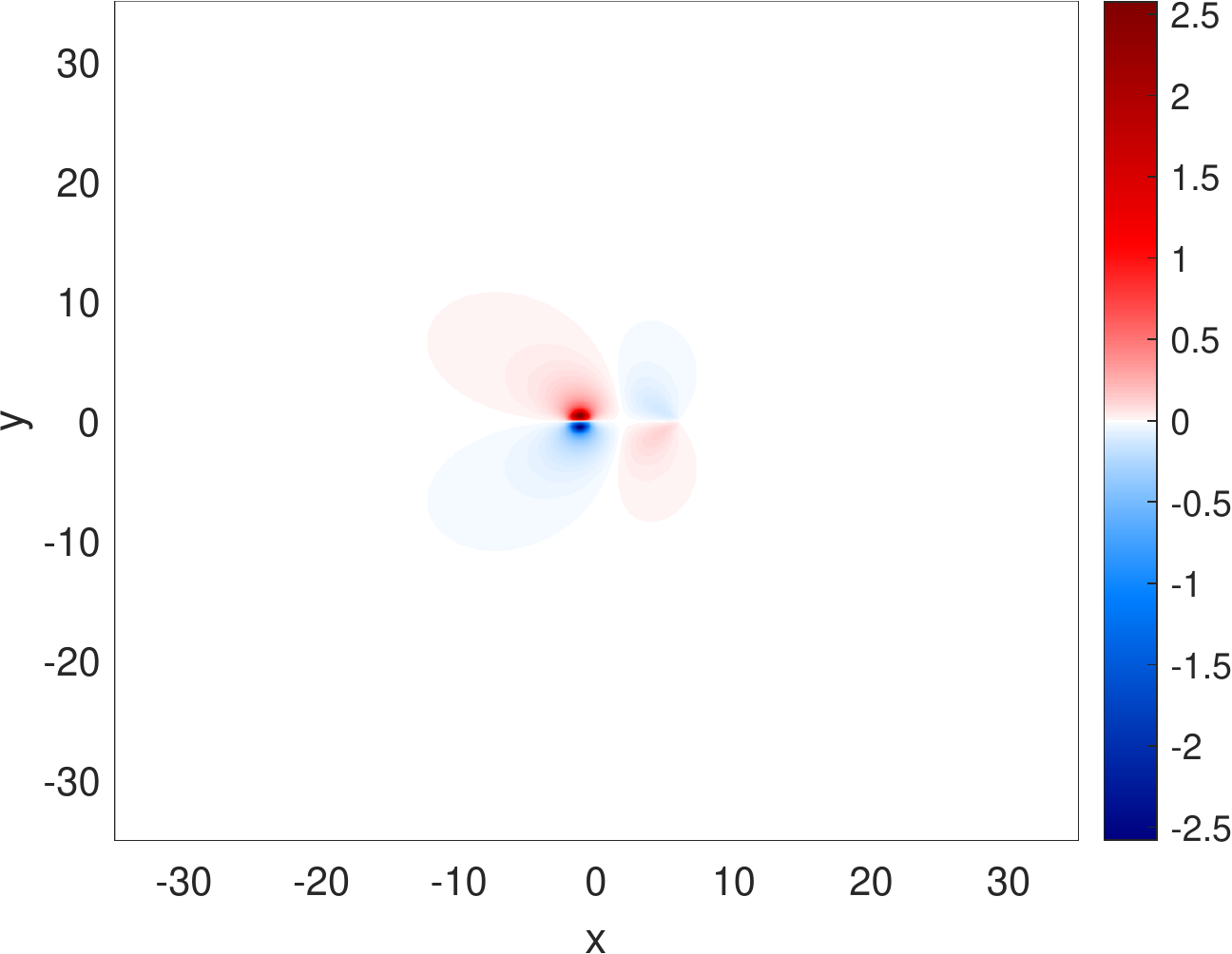}
	\caption{}
	\end{subfigure}
	\begin{subfigure}[b]{0.49\textwidth}
	\centering
	\includegraphics[trim={0cm 0cm 0cm 0cm},clip,width=\textwidth]{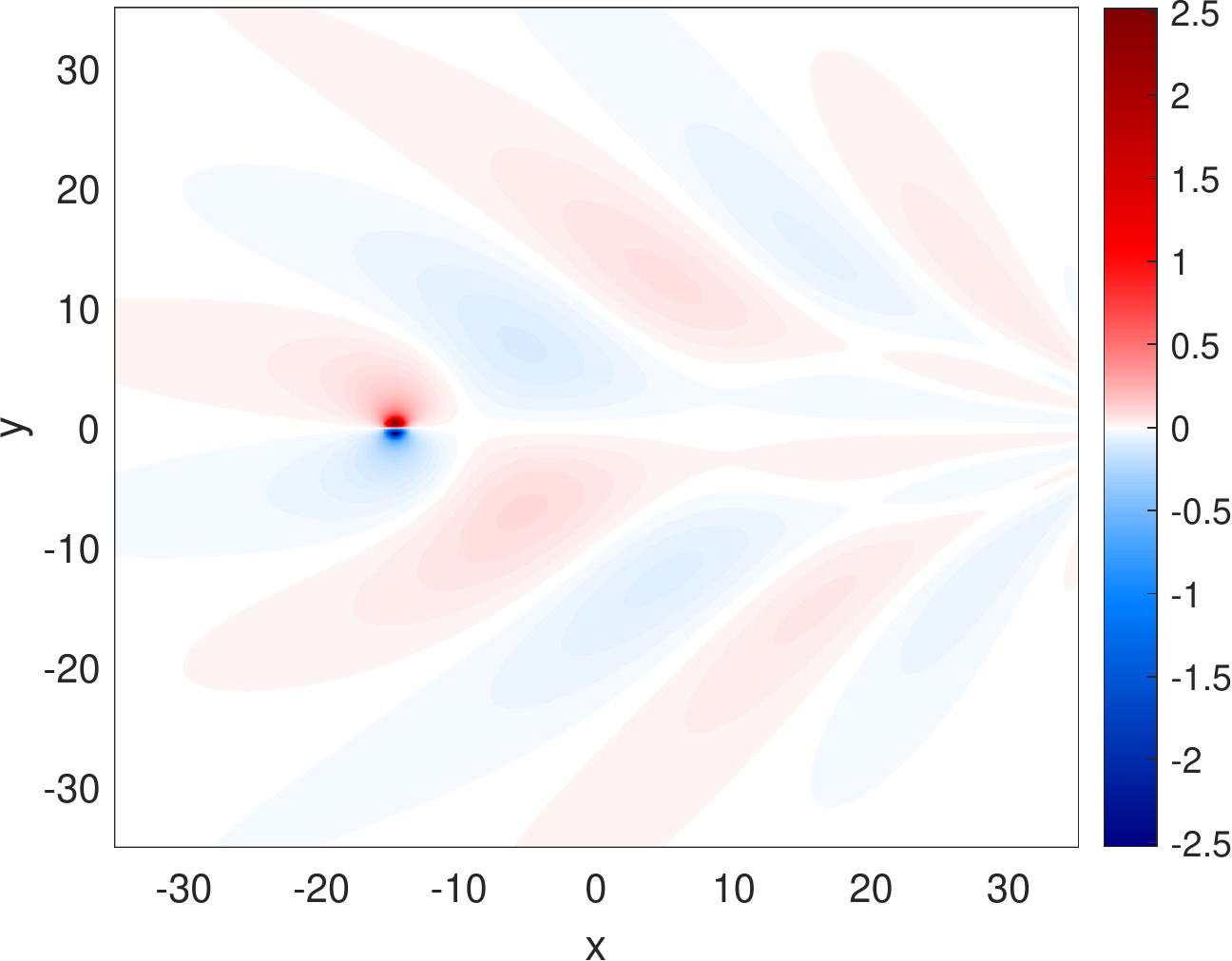}
	\caption{}
	\end{subfigure}
	\caption{The streamfunction $\psi$ as a function of $(x,y)$ for $\lambda = 0.4$ at times (a) $t = 5$ and (b) $t = 40$.}
    \label{fig:slices1}
\end{figure}

\begin{figure}
    \centering
	\begin{subfigure}[b]{0.49\textwidth}
	\centering
	\includegraphics[trim={0cm 0cm 0cm 0cm},clip,width=\textwidth]{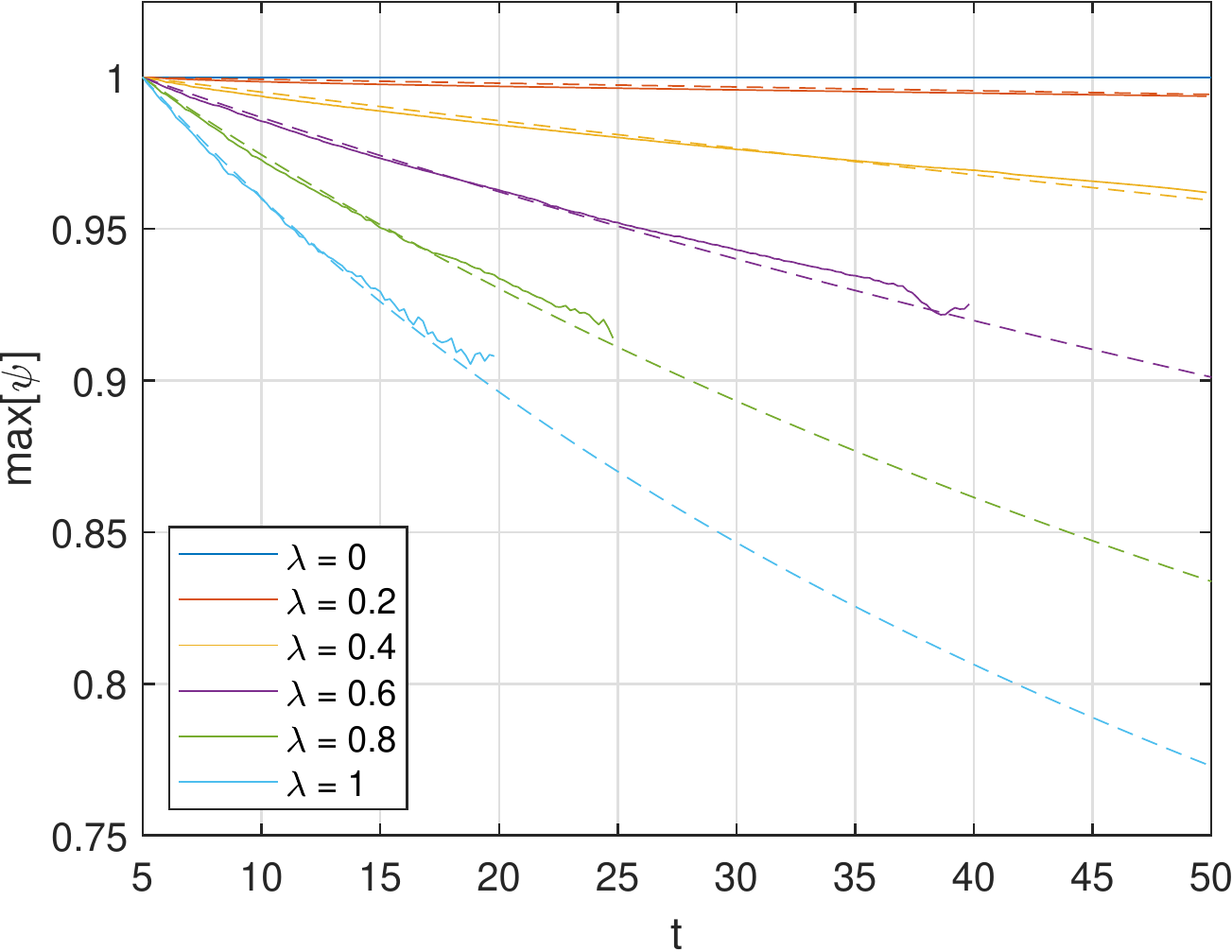}
	\caption{}
	\end{subfigure}
	\begin{subfigure}[b]{0.49\textwidth}
	\centering
	\includegraphics[trim={0cm 0cm 0cm 0cm},clip,width=\textwidth]{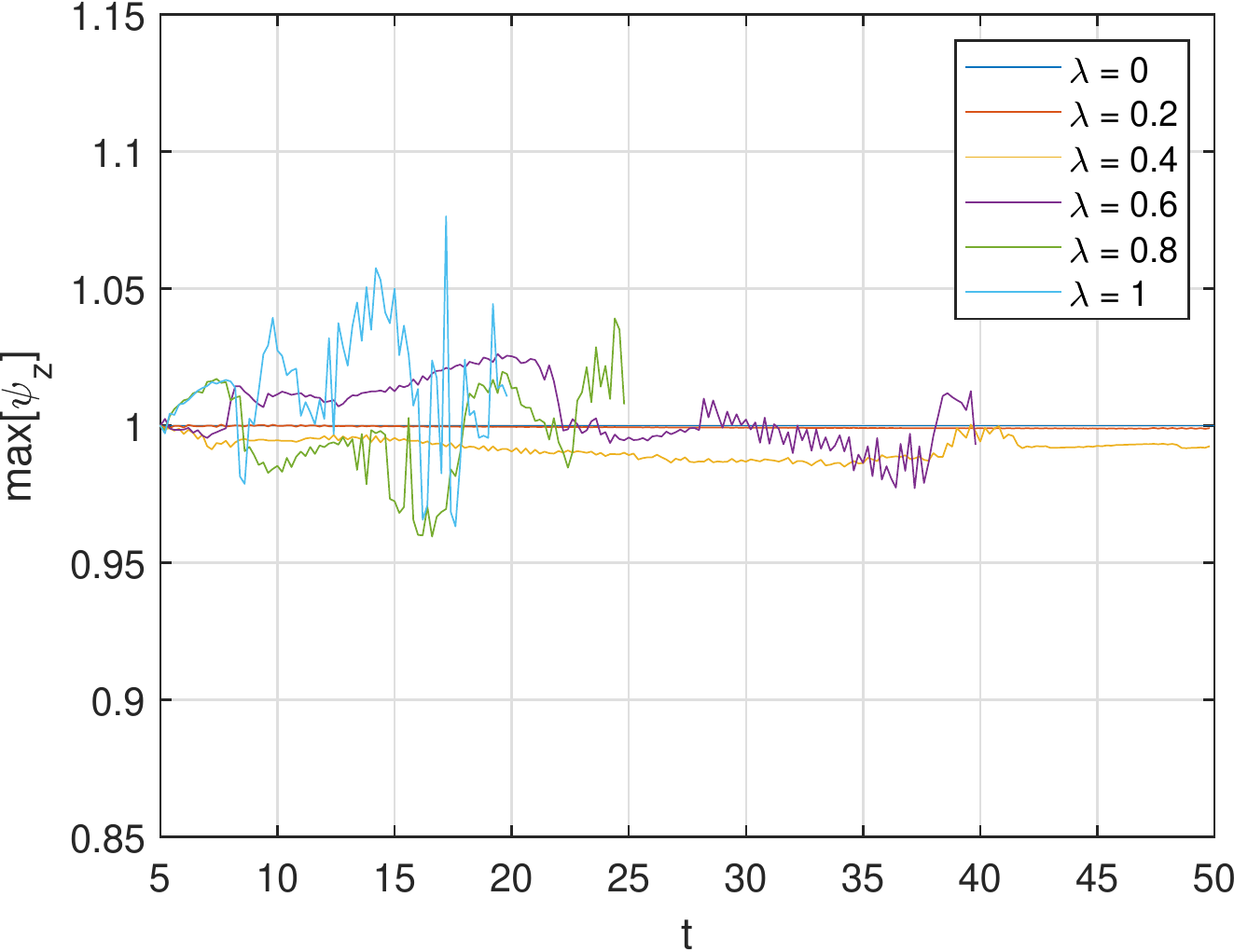}
	\caption{}
	\end{subfigure}
	\caption{Plots of (a) $\max [\psi]$ and (b) $\max [\pderline{\psi}{z}]$ from numerical simulations for $\lambda \in \{0, 0.2, 0.4, 0.6, 0.8, 1\}$. Results are shown as functions of time, $t$, from time $t_0 = 5$ onwards and all curves are normalised by their initial value. Dashed lines denote the asymptotic predictions of \cref{eq:decay_pred}.}
    \label{fig:data_1}
\end{figure}

\cref{fig:slices1} shows $\psi$ from our numerical simulations at times $t = t_0 = 5$ and $t = 40$ for slope $\lambda = 0.4$. The formation of a wave wake can clearly be seen. Results are shown in coordinates moving with speed $C = -1$. Since the vortex speed, $U$, changes slightly during the initial adjustment phase, the vortex does not remain at the origin throughout the evolution. \cref{fig:data_1} shows the maximum values of $\psi$ and $\pderline{\psi}{z}$ as functions of time for each simulation run. Our asymptotic decay predictions from \cref{eq:decay_pred} are shown as dashed lines and found to give accurate predictions for the decay, even for the cases where $\lambda$ is order one. The value of $\max[\pderline{\psi}{z}]$ is found to be well conserved throughout the evolution---varying by less than $3\%$ for most cases---justifying the assumptions made in \cref{sec:decay_pred}. The curves for $\lambda \geq 0.6$ in \cref{fig:data_1} are shown over the limited interval $t \in [t_0, t_\lambda]$ where $t_\lambda = 40, 25, 20$ respectively for $\lambda = 0.6, 0.8, 1.0$. Outside these intervals, the dipolar vortices break down and the assumption of self-similar vortex structure is no longer valid. This breakdown is explored in \cref{sec:breakdown}.

\section{Vortex breakdown}
\label{sec:breakdown}

Numerical solutions reveal that over long times, decaying dipolar vortices with $\mu < 0$ will break-down into a pair of monopolar vortices. When this breakdown occurs, the structures cease to move purely in the East-West direction and instead follow complex trajectories. This breakdown occurs sooner for steeper slopes i.e. sooner for larger values of $\lambda$ and $|\mu|$.

\begin{figure}
    \centering
	\begin{subfigure}[b]{0.49\textwidth}
	\centering
	\includegraphics[trim={0cm 0cm 0cm 0cm},clip,width=\textwidth]{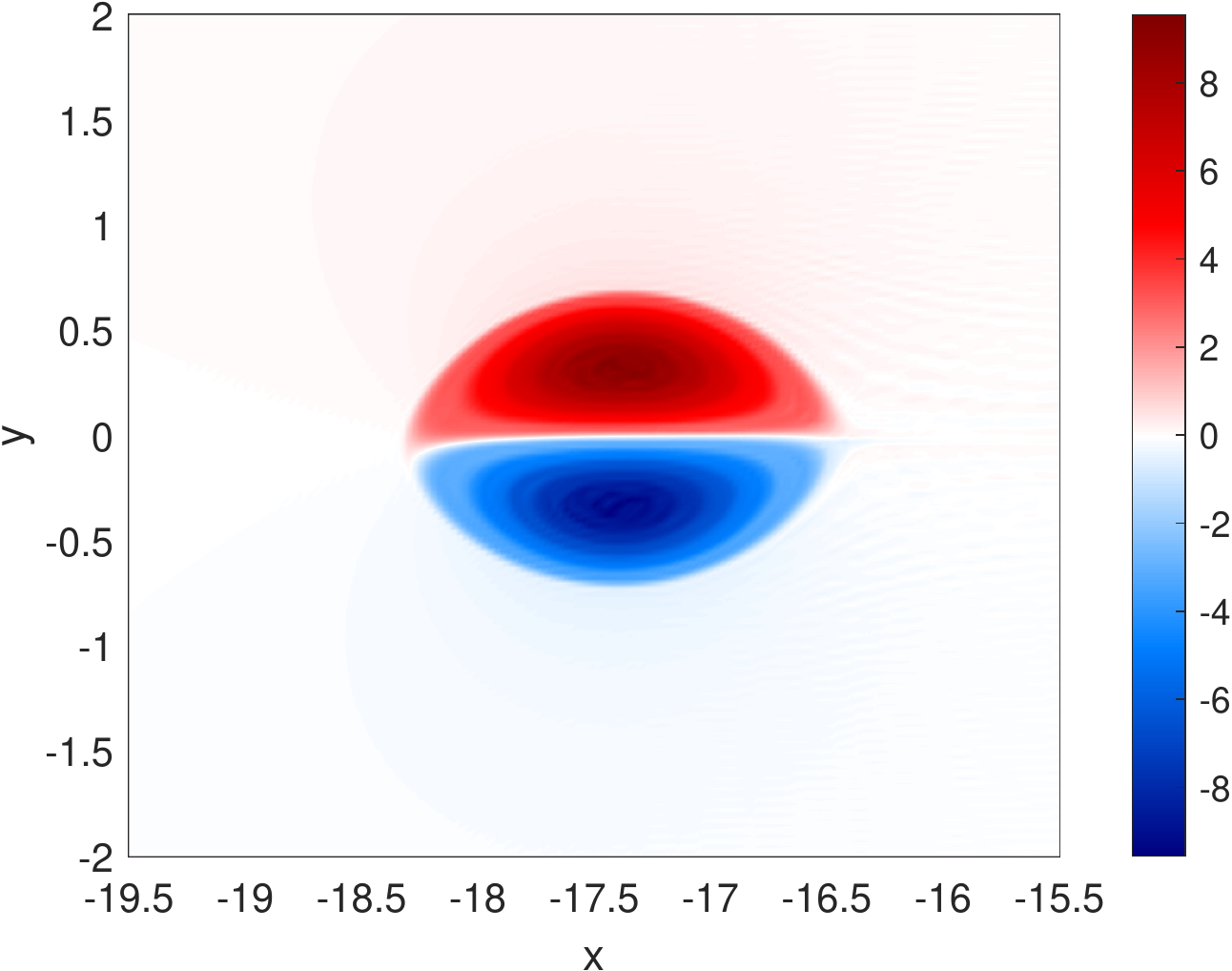}
	\caption{}
	\end{subfigure}
	\begin{subfigure}[b]{0.49\textwidth}
	\centering
	\includegraphics[trim={0cm 0cm 0cm 0cm},clip,width=\textwidth]{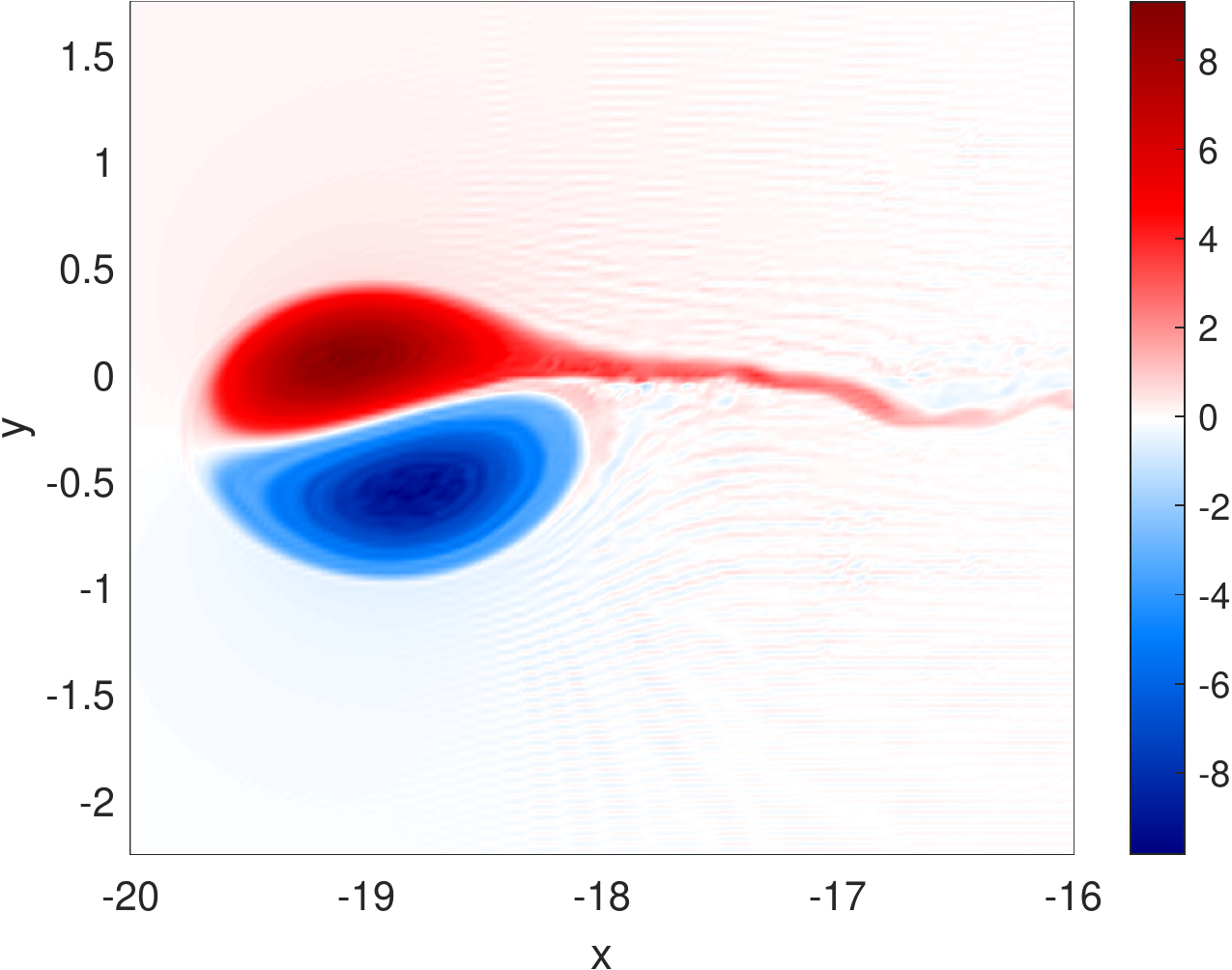}
	\caption{}
	\end{subfigure}
		\begin{subfigure}[b]{0.49\textwidth}
	\centering
	\includegraphics[trim={0cm 0cm 0cm 0cm},clip,width=\textwidth]{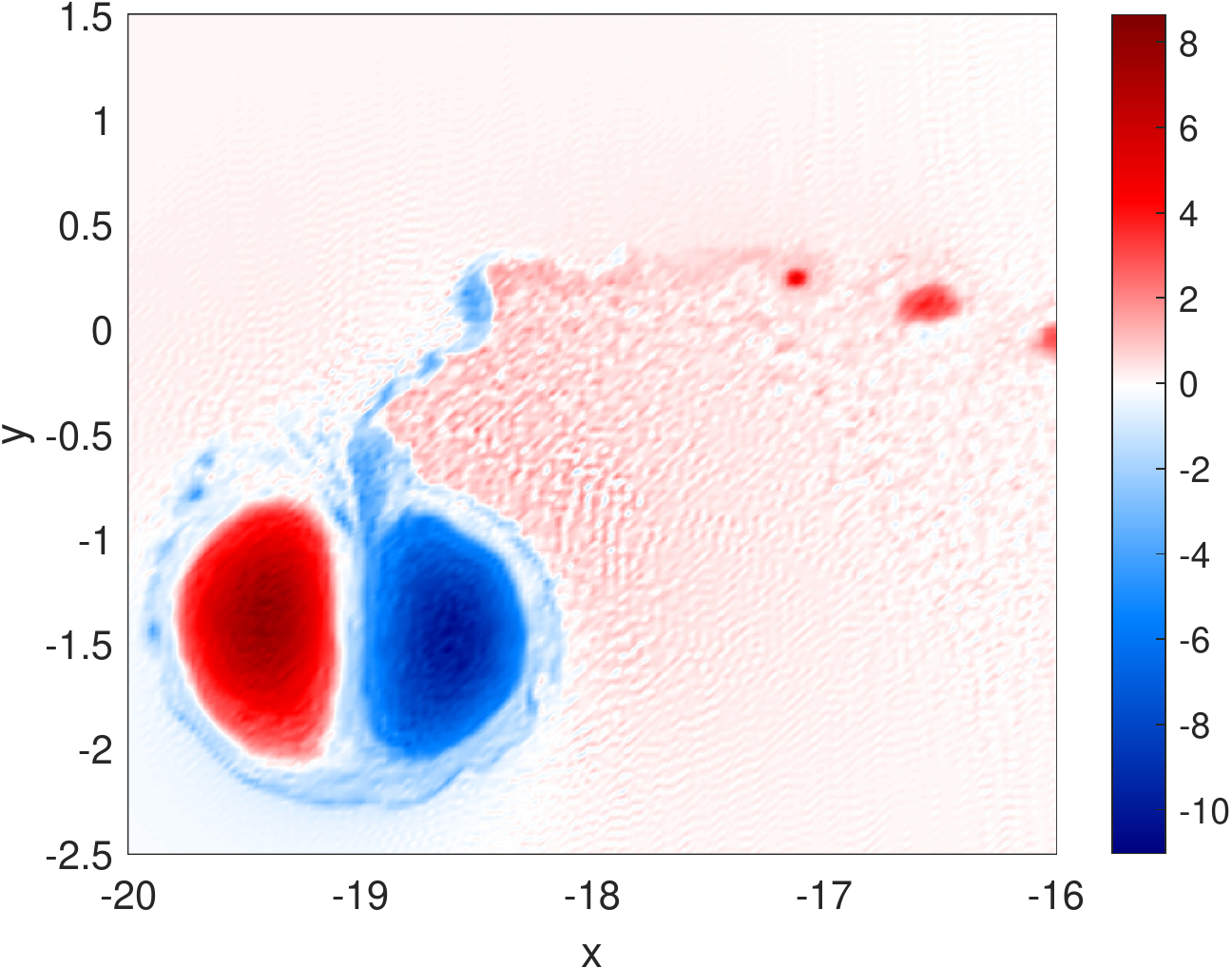}
	\caption{}
	\end{subfigure}
	\begin{subfigure}[b]{0.49\textwidth}
	\centering
	\includegraphics[trim={0cm 0cm 0cm 0cm},clip,width=\textwidth]{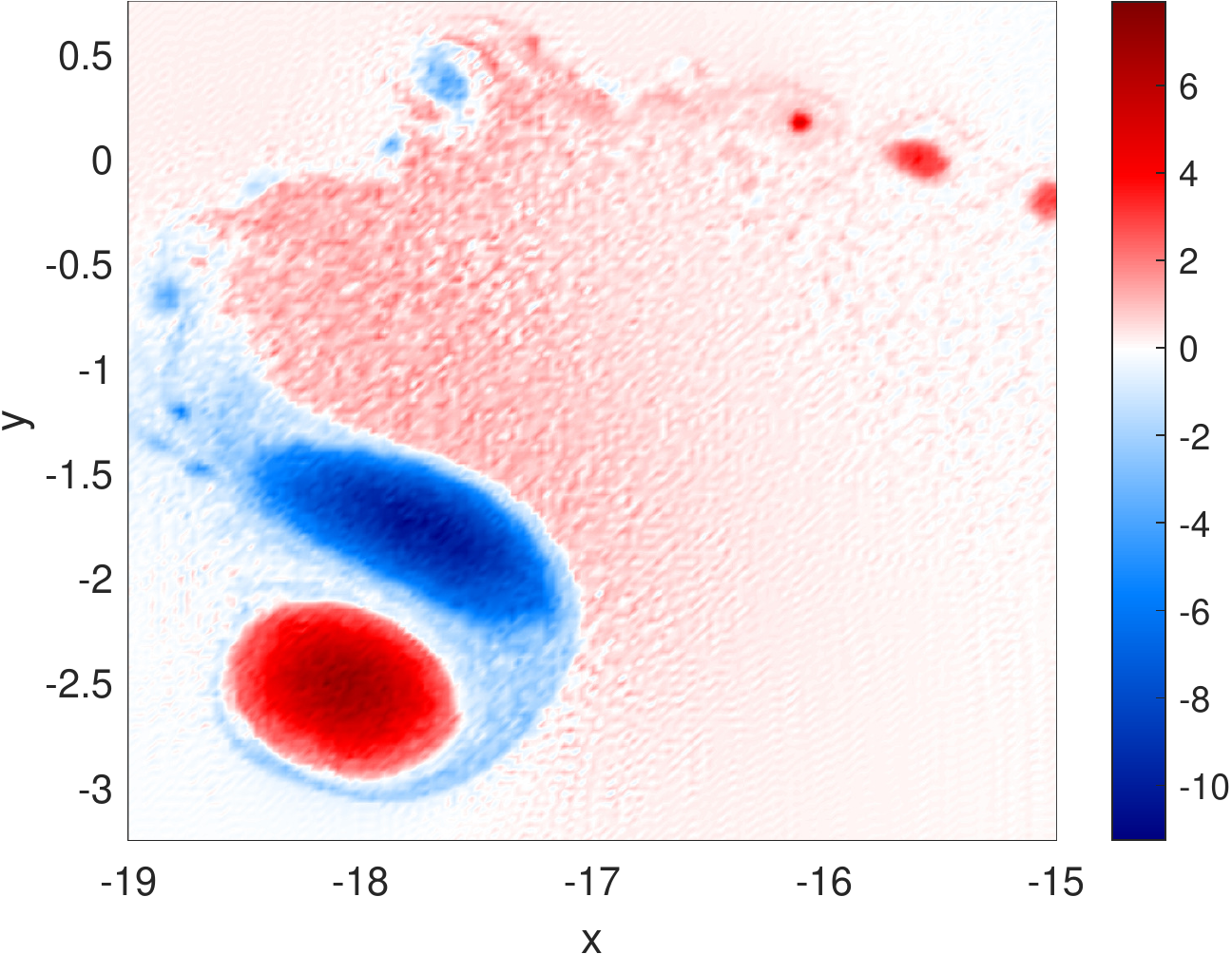}
	\caption{}
	\end{subfigure}
		\begin{subfigure}[b]{0.49\textwidth}
	\centering
	\includegraphics[trim={0cm 0cm 0cm 0cm},clip,width=\textwidth]{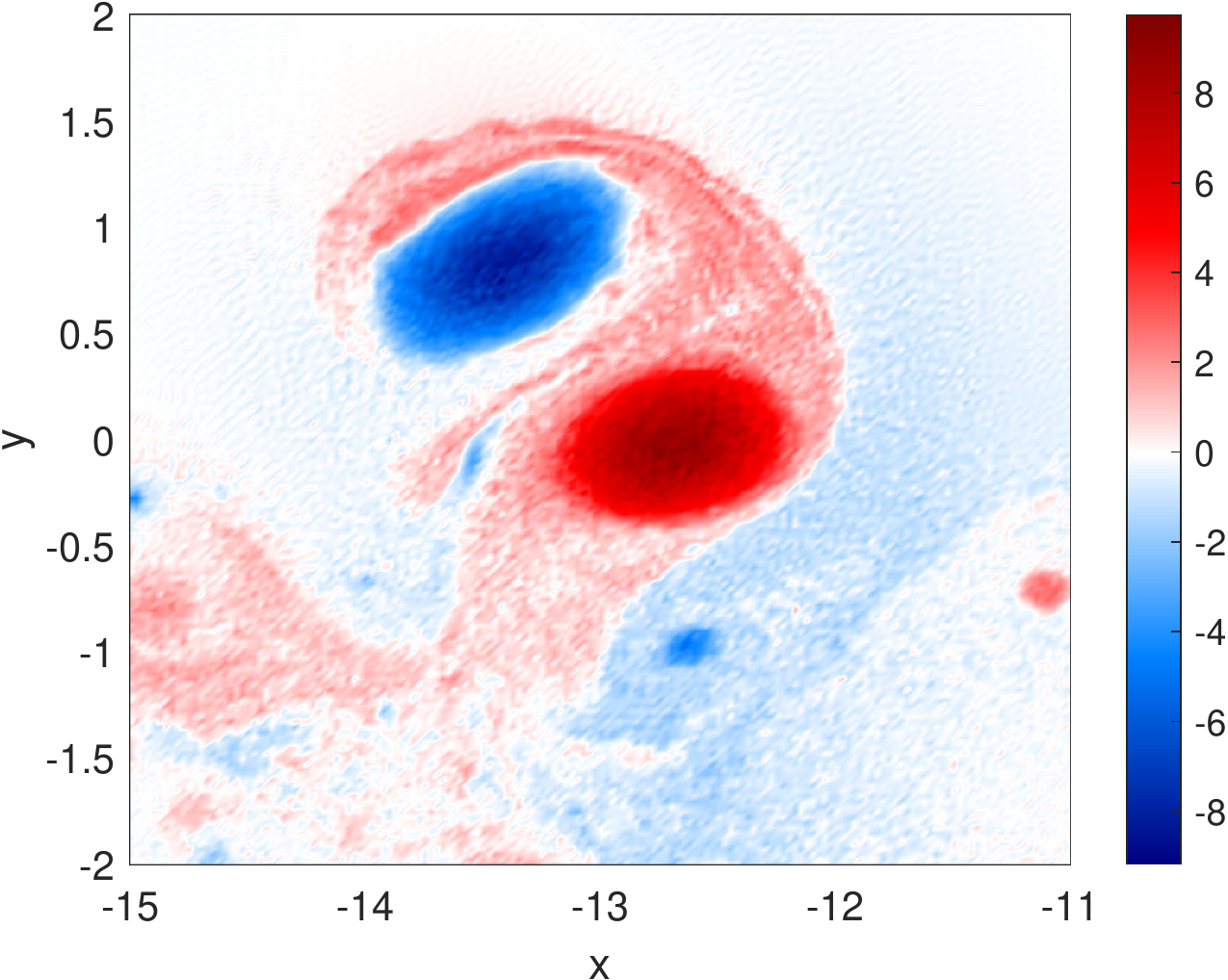}
	\caption{}
	\end{subfigure}
	\begin{subfigure}[b]{0.49\textwidth}
	\centering
	\includegraphics[trim={0cm 0cm 0cm 0cm},clip,width=\textwidth]{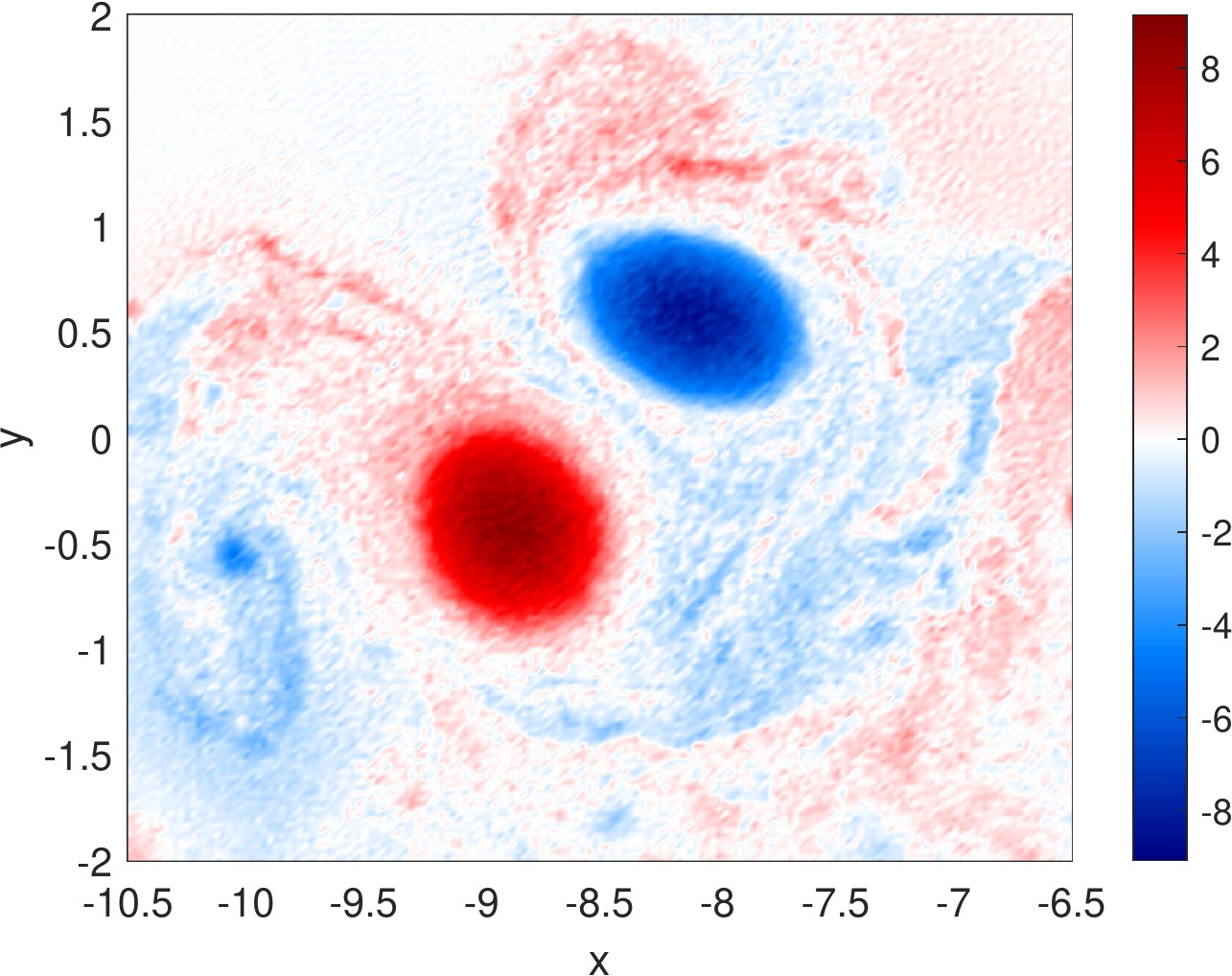}
	\caption{}
	\end{subfigure}
	\caption{Plots of the surface buoyancy, $\pderline{\psi}{z}$, for $\lambda = 0.8$ and (a) $t = 25$, (b) $t = 27$, (c) $t = 29$, (d) $t = 30$, (e) $t = 35$ and (f) $t = 40$. The breakdown of a dipolar vortex into two monopolar vortices can be clearly seen.}
    \label{fig:breakdown}
\end{figure}

Examination of simulation results suggests that the breakdown occurs through the development of waves inside the vortex. These waves lead to an asymmetry which causes the two halves of the dipole to split. The wave amplitude is larger for larger values of $|\mu|$ hence the breakdown occurs sooner. \cref{fig:breakdown} shows the breakdown of a dipolar vortex for $(U,a,\lambda) = (-1,1,0.8)$ which occurs from around $t_\lambda = 25$. The breakdown is most clearly seen in the buoyancy field, $\pderline{\psi}{z}$, hence we plot buoyancy rather than streamfunction, $\psi$. We observe that after an asymmetry develops in the dipolar structure, streamers of buoyancy (and vorticity) are stripped away from each side. These streamers later roll up into lines of small monopolar vortices which interact with each other through a series of merging and splitting events as predicted by the theory of SQG turbulence \citep{HELD_ET_AL_1995}. As the initial asymmetry results in the two halves having a different vortex strength, they cease to move together in a straight line. Instead, the modified vortex pair deflects South before following a complicated path as the vortex strengths changes due to subsequent interactions. As expected, waves are still generated post-breakdown if there is a significant disturbance moving Westwards. A supplementary movie (Movie\_1.mp4) shows the full evolution of the modon for the case of $\lambda = 0.8$ depicted in \cref{fig:breakdown}.

\citet{Snyder_et_al_2007} considered the evolution of SQG modons in a primitive equation model and observed that waves could form within the modon. These inertia-gravity waves were found to result from both initial adjustment and imbalance and were stronger for larger Rossby numbers. In our case, we are considering a balanced model and, as such, expect that waves occur through a combination of initial adjustment and topographic effects. A small disturbance advected in the up-slope, $y$, direction by the interior flow can increase in vorticity due to the conservation of PV. This magnification of small disturbances may in turn lead to vortex breakdown.

\section{Stabilisation by a coastal boundary}

By the image effect, we would expect a monopolar vortex moving along a coastal boundary to behave in the same way as a dipolar modon and be described by the same solutions as discussed in \cref{sec:setup,sec:steady_sol,sec:decay_sol}. In particular, we expect steady solutions for retrograde vortices and a decay rate for prograde vortices which follows \cref{eq:decay_pred} in the limit of small $\lambda$.

\begin{figure}
    \centering
    \begin{subfigure}[b]{0.49\textwidth}
	\centering
	\includegraphics[trim={0cm 0cm 0cm 0cm},clip,width=\textwidth]{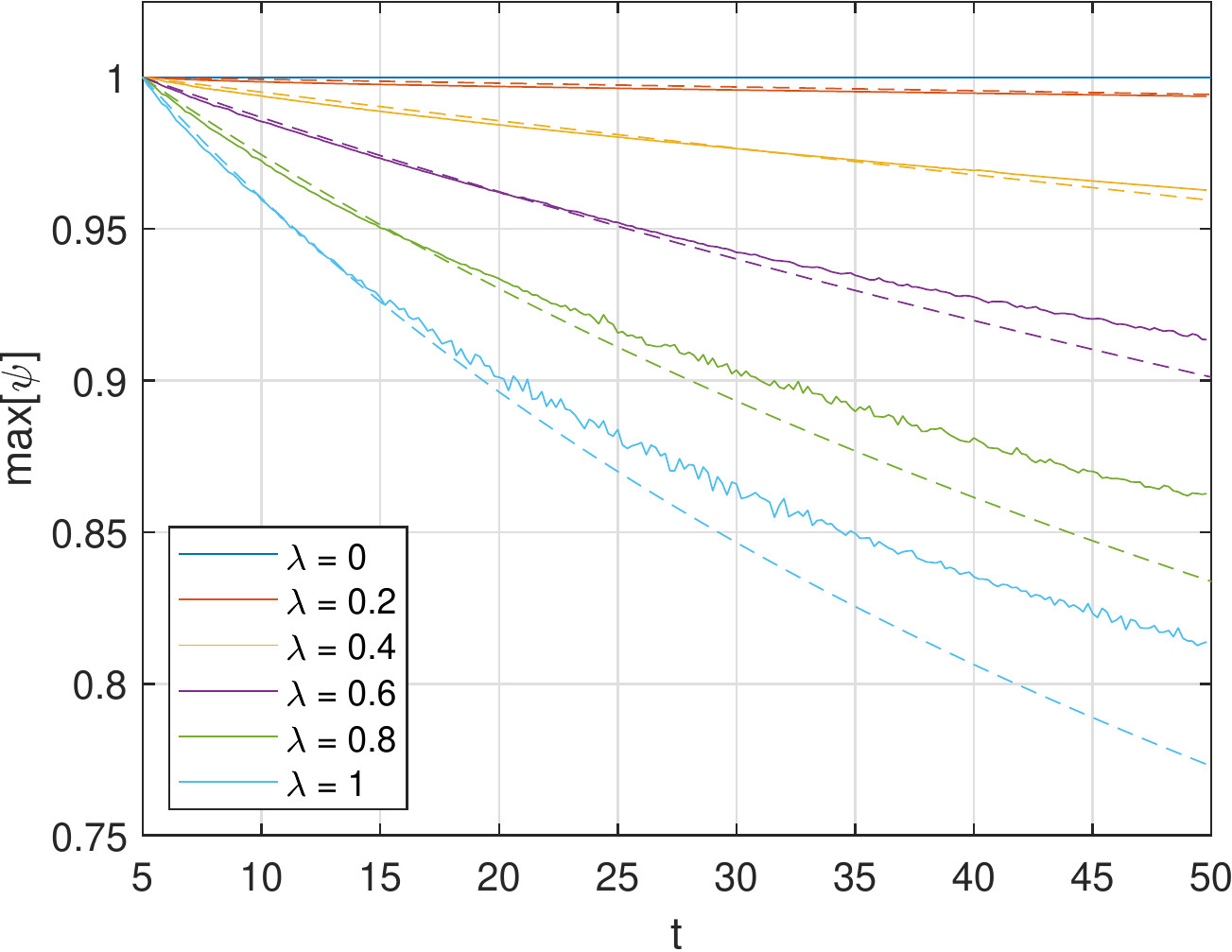}
	\caption{}
	\end{subfigure}
	\begin{subfigure}[b]{0.49\textwidth}
	\centering
	\includegraphics[trim={0cm 0cm 0cm 0cm},clip,width=\textwidth]{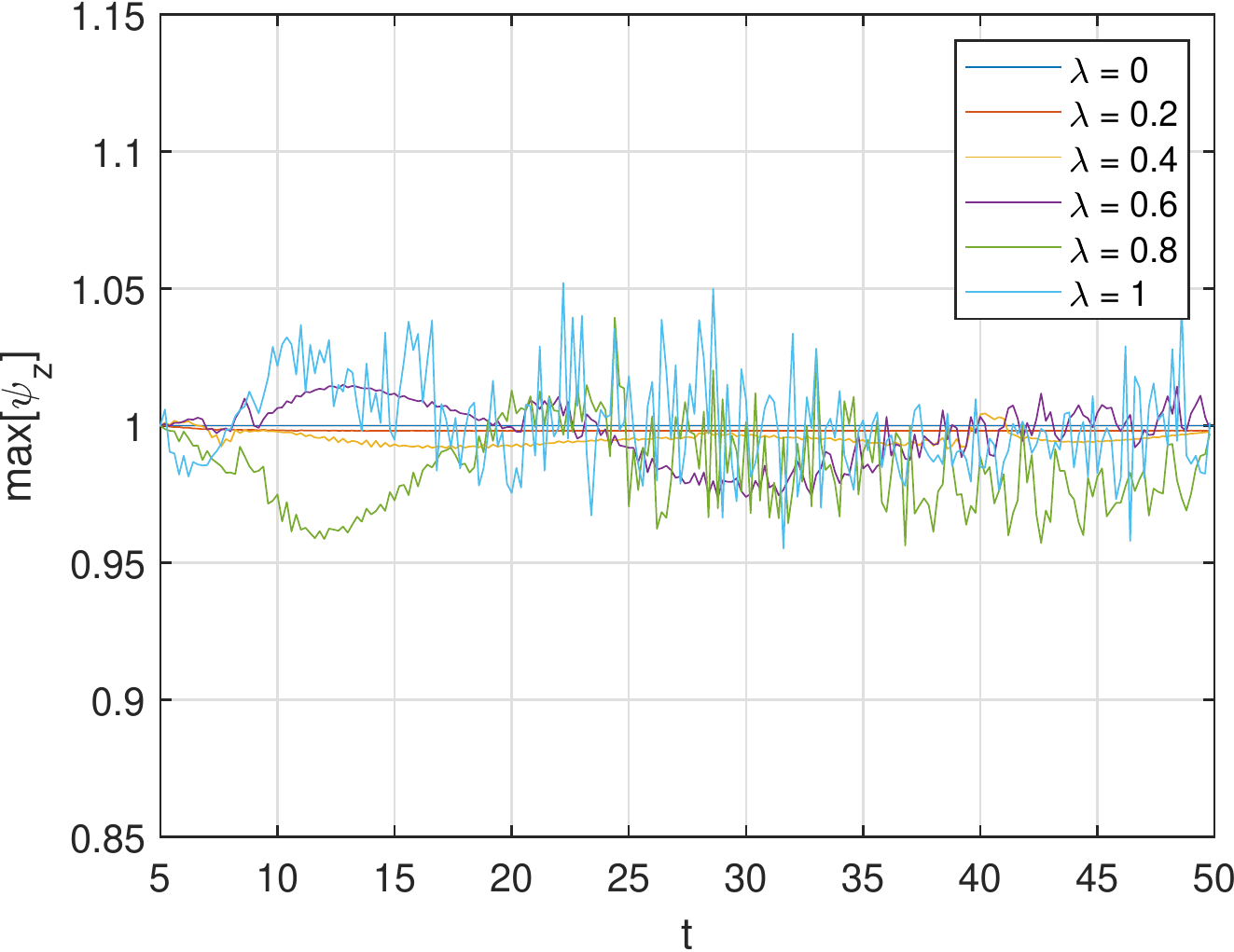}
	\caption{}
	\end{subfigure}
    \caption{As \cref{fig:data_1} for a monopolar vortex moving along a coastal boundary.}
    \label{fig:wall_decay}
\end{figure}

\begin{figure}
    \centering
	\begin{subfigure}[b]{0.49\textwidth}
	\centering
	\includegraphics[trim={0cm 0cm 0cm 0cm},clip,width=\textwidth]{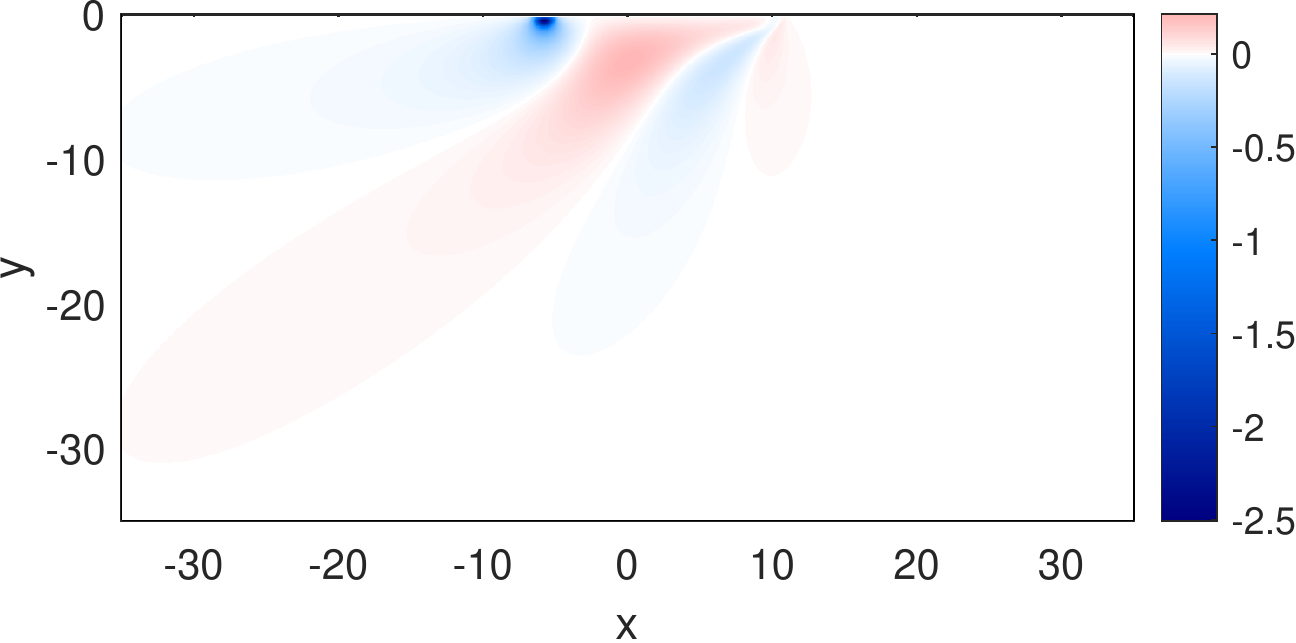}
	\caption{}
	\end{subfigure}
	\begin{subfigure}[b]{0.49\textwidth}
	\centering
	\includegraphics[trim={0cm 0cm 0cm 0cm},clip,width=\textwidth]{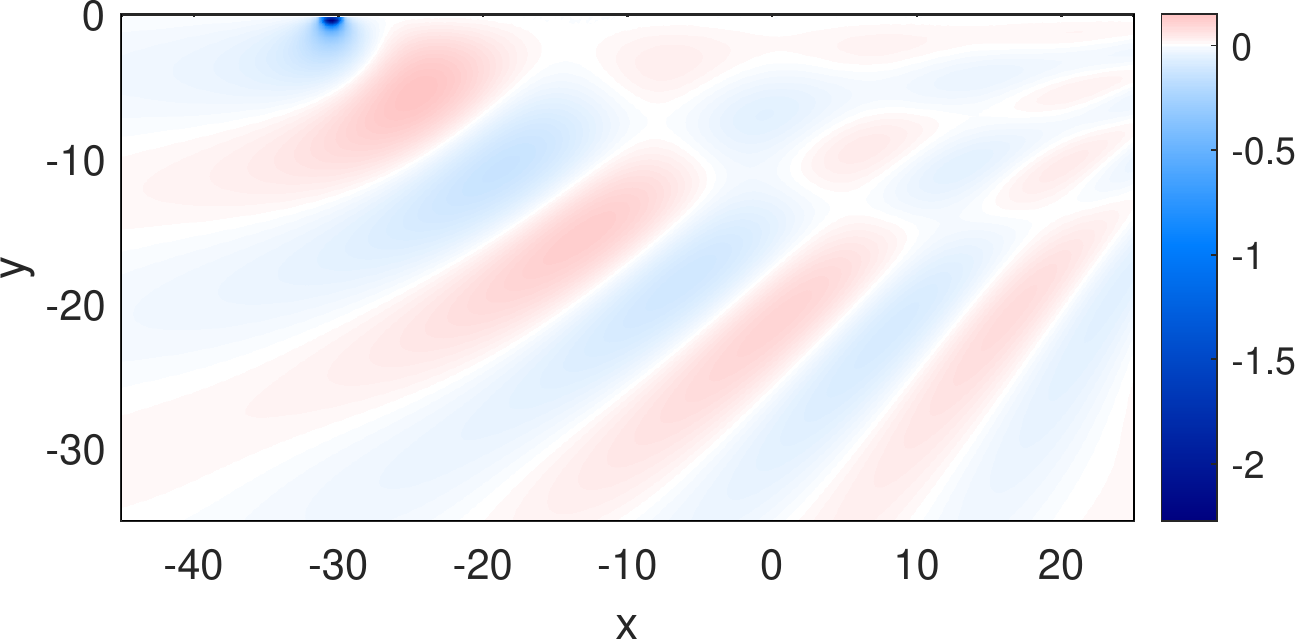}
	\caption{}
	\end{subfigure}
		\begin{subfigure}[b]{0.49\textwidth}
	\centering
	\includegraphics[trim={0cm 0cm 0cm 0cm},clip,width=\textwidth]{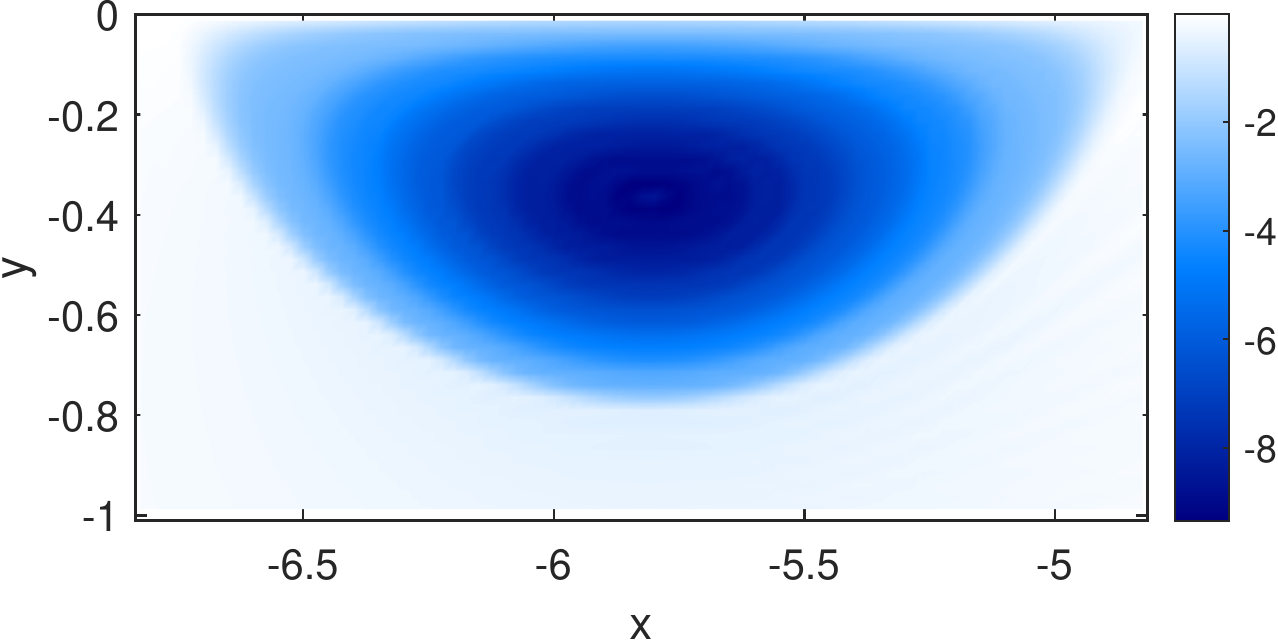}
	\caption{}
	\end{subfigure}
	\begin{subfigure}[b]{0.49\textwidth}
	\centering
	\includegraphics[trim={0cm 0cm 0cm 0cm},clip,width=\textwidth]{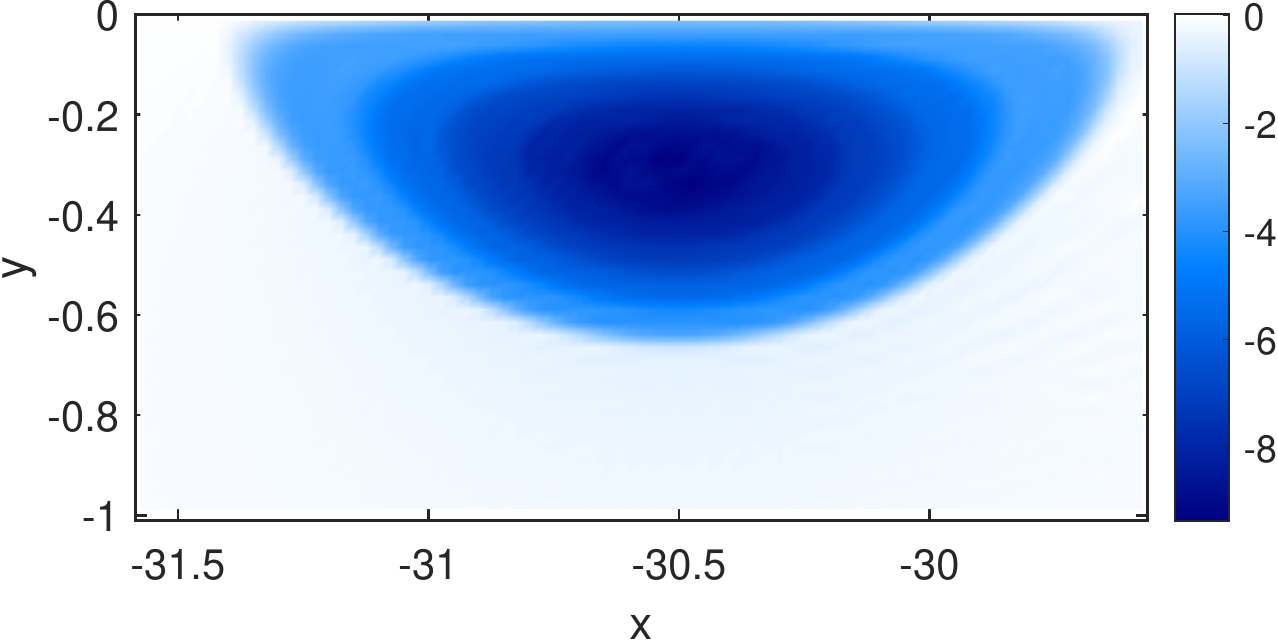}
	\caption{}
	\end{subfigure}
	\caption{The time evolution of a monopolar vortex on a wall for $\lambda = 0.8$. We plot the surface streamfunction, $\psi$, for (a) $t = 10$ and (b) $t = 40$ and the surface buoyancy, $\pderline{\psi}{z}$, for (c) $t = 10$ and (d) $t = 40$.}
    \label{fig:wall_evol}
\end{figure}

To test our decay prediction, we rerun our Dedalus simulations on the half domain $(x,y) \in [-51.2,51.2]\times[-51.2,0]$ using a Sin basis in the $y$ direction to enforce the condition that $\psi = 0$ along the wall, $y = 0$. All other parameters and analysis techniques remain unchanged. \cref{fig:wall_decay} shows the maximum values of $\psi$ and $\pderline{\psi}{z}$ as functions of time for each simulation run. Our asymptotic prediction from \cref{eq:decay_pred} is included as a dotted line and found to closely match the observed decay, even for order $1$ values of $\lambda$. \cref{fig:wall_evol} shows the surface streamfunction and buoyancy for the case of $\lambda = 0.8$ at two different times. The formation of the wave field can be clearly seen in the plots of $\psi$ while the close-up plots of $\partial{\psi}{z}$ show the decrease in vortex size at later times. We observe that at late times and larger values of $\lambda$, the vortex shape becomes elliptical with a shorter semi-radius in the $y$ direction. This will reduce the validity of the assumption that the vortex remains self-similar throughout the evolution and may partially explain why the decay prediction is less valid in these cases. A supplementary movie (Movie\_2.mp4) shows the full evolution of the coastal vortex for the case of $\lambda = 0.8$ depicted in \cref{fig:wall_evol}.

A major difference observed between the dipolar vortices and the coastal monopoles is that the vortex breakdown discussed in \cref{sec:breakdown} does not occur during the monopole evolution. This is likely a consequence of the enforced symmetry of the image vortex across the line $y = 0$ and suggests that the breakdown of the dipoles occurs due to the growth of an asymmetry between the regions of positive and negative vorticity.

\section{Discussion and conclusions}
\label{sec:discuss}
We have studied the evolution of a propagating modon on sloping topography using a combination of numerical and analytical approaches. Since Rossby waves propagate along the slope in one direction only, there are two regimes of interest: retrograde motion where the modon propagates oppositely to the waves and prograde motion where the waves and modon propagate in the same direction.

In steady retrograde motion no wave wake is generated as there are no waves with phase speed matching the speed of the vortex. Steady vortex solutions exist which can be found by expanding the vortex solution in terms of Zernike radial functions. The coefficients of the Zernike functions, and the internal wavenumber of the modon $K$, follow by solving a simple linear algebraic eigenvalue problem. Increasing the slope gradient is found to increase the velocity of fluid within the vortex and create regions surrounding the vortex core where the buoyancy is oppositely signed to the closest peak.

In prograde motion waves are generated with a phase speed matching the speed of the vortex. As no energy enters the system from upstream, a unique solution can be found in which a wave wake is generated behind the vortex. This wake removes energy from the vortex in the form of a radial energy flux from the vortex into the wave field. Under the assumption of weak slope, we derive predictions for the decay of the vortex radius with time. In contrast with the similar problem of a one-layer QG dipole on a beta-plane \citep{FLIERLHAINES94,crowe_johnson_2020}, the vortex speed is predicted to remain unchanged throughout the decay. Our asymptotic predictions are tested numerically and found to closely match simulation results. In particular, we found that decay predictions remain accurate even for order one values of the slope parameter. Over long times we observe that modons break down as asymmetries develop between the two halves of the dipole. These asymmetries lead to the formation of two coupled monopolar vortices and are predicted to occur due to wave-like oscillations within the modon.

Solutions corresponding to a monopolar vortex moving along a coastal boundary can be found using the method of images. These solutions are identical to our dipolar solutions over the reduced domain and follow a similar vortex decay in the prograde case. Over long times, the vortex breakdown observed for prograde modons does not occur, likely as a result of the enforced symmetry of the wall.

Due to the widespread nature of dipolar vortices in both the ocean and atmosphere, our results and predictions may be applicable to a range of phenomena. For example, the mathematical system we study is identical to the case of a surface modon moving on a density front in the along-front direction. Therefore, our results may be relevant to propagating modons in the Gulf Stream and Southern Ocean, both very energetic regions of the ocean containing a high density of modons and fronts \citep{NiZWH20,ORSI_1995}. Similarly, modons have been used as simple models for various atmospheric phenomena, such as localised jets at the tropopause and atmospheric blocks \citep{MurakiS07}, so our results may be relevant for modelling these flows or understanding their generation of Rossby waves.\\

\noindent{\bf Funding\bf{.}} This work was funded by the UK Natural Environment Research Council under grant number NE/S009922/1.\\

\noindent{\bf Declaration of Interests\bf{.}} The authors report no conflict of interest.

\appendix

\section{Evaluation of Bessel function integrals}
\label{sec:bessel_eval}

Here we discuss the numerical approach used to evaluate the oscillatory integrals in \cref{eq:A_def,eq:B_def}. Consider an integral of the form
\begin{equation}
F_{mn} = \int_0^\infty f(x) \J_m(x) \J_n(x) \dint x,
\end{equation}
where $f = O(1/x)$ as $x\to \infty$ so the integral converges. Using the relation
\begin{equation}
\J_m(x) = \frac{1}{2}\left[ \H_m^{(1)}(x)+\H_m^{(2)}(x)\right],
\end{equation}
where $\H_m^{(1)}$ and $\H_m^{(2)}$ are Hankel functions of the first and second kind respectively, we may write
\begin{equation}
\label{eq:F_mn_2}
F_{mn} = \int_0^d f(x) \J_m(x) \J_n(x) \dint x + \int_d^\infty \frac{f(x)}{4}\left[ \mathcal{H}_{mn}^{1,1}(x)+\mathcal{H}_{mn}^{1,2}(x)+\mathcal{H}_{mn}^{2,1}(x)+\mathcal{H}_{mn}^{2,2}(x) \right] \dint x,
\end{equation}
for
\begin{equation}
\mathcal{H}_{mn}^{i,j}(x) = \H_m^{(i)}(x)\H_m^{(j)}(x).
\end{equation}
Here we have split $F_{mn}$ at $x = d$; the first part can be accurately determined numerically, while the second part is strongly oscillatory and hence is difficult to evaluate directly. We now consider the terms in the second integral of \cref{eq:F_mn_2} separately.

Consider first
\begin{equation}
I^{(1)}_{mn} = \frac{1}{4} \int_d^\infty f(x)\, \mathcal{H}^{1,1}_{mn}(x) \dint x.
\end{equation}
This integral is oscillatory though the integration contour may be deformed into the upper half-plane where the integrand decays exponentially. We use the contour $z = d+iy$ for $y>0$ and change variables to $y$ to obtain
\begin{equation}
I^{(1)}_{mn} = \frac{i}{4} \int_0^\infty f(d+iy)\, \mathcal{H}^{1,1}_{mn}(d+iy) \dint y.
\end{equation}
Similarly the integral
\begin{equation}
I^{(2)}_{mn} = \frac{1}{4} \int_d^\infty f(x)\, \mathcal{H}^{2,2}_{mn}(x) \dint x,
\end{equation}
has an integrand which decays exponentially in the lower half-plane and may be deformed to get
\begin{equation}
I^{(2)}_{mn} = -\frac{i}{4} \int_d^\infty f(d-iy)\, \mathcal{H}^{2,2}_{mn}(d-iy) \dint y.
\end{equation}
Finally, the cross-term integral from \cref{eq:F_mn_2}
\begin{equation}
I^{(3)}_{mn} = \frac{1}{4} \int_d^\infty f(x) \left[ \mathcal{H}^{1,2}_{mn}(x) + \mathcal{H}^{2,1}_{mn}(x) \right]\dint x,
\end{equation}
is non-oscillatory and decays as a power law for large $x$ so may be accurately evaluated numerically.

Our final result is then determined as
\begin{equation}
F_{mn} = \int_0^d f(x) \J_m(x) \J_n(x) \dint x + I^{(1)}_{mn} + I^{(2)}_{mn} + I^{(3)}_{mn},
\end{equation}
and a value of $d = 10^3$ is used in our calculations. The function $f$ is given by $f(x) = 1/(x+\mu)$ and $f(x) = 1/(x^2+\mu x)$ when calculating $A_{mn}$ and $B_{mn}$ respectively.  We note that for $\mu = 0$ these integrals may instead be evaluated analytically \citep{JohnsonC22}.

\section{Leading order energy and momentum}

The vortex energy and ($x$) momentum are given by
\begin{equation}
\label{eq:E_vol}
E = \frac{1}{2}\int_0^\infty\!\!\! \int_{-\infty}^\infty\! \int_{-\infty}^\infty\!\! |\nabla \psi|^2 \dint x \dint y \dint z = -\frac{1}{2}\int_{-\infty}^\infty\! \int_{-\infty}^\infty\! \left[\psi \pder{\psi}{z}\right]_{z=0} \!\dint x \dint y,
\end{equation}
and
\begin{equation}
m = -\int_0^\infty\!\!\! \int_{-\infty}^\infty\! \int_{-\infty}^\infty\! \pder{\psi}{y} \dint x \dint y \dint z = \int_{-\infty}^\infty\! \int_{-\infty}^\infty\! \left[y \pder{\psi}{z}\right]_{z=0} \!\dint x \dint y.
\end{equation}
Therefore, to leading order in small $\mu$, the vortex momentum is given by
\begin{equation}
\label{eq:m_def}
m = - Ua^3 \sum_{n=0}^\infty a_n \int_0^{2\pi} \!\!\! \sin^2\theta  \dint \theta \int_0^1 \! s^2 \R_n(s) \dint s = -\frac{a_0 \pi U a^3}{4}, 
\end{equation}
which we note is positive as $-a_0/4 \approx 1.55156$. The leading order vortex energy may be similarly calculated as
\begin{equation}
E = \frac{1}{2}\int_{0}^{2\pi}\!\!\! \int_{0}^a\! \left[\pder{\psi}{z} \left( \frac{a}{K}\pder{\psi}{z} + Uy \right)\right]_{z=0} \!\!\! r \dint r \dint \theta = \frac{a}{2K}\int_{0}^{2\pi}\!\!\! \int_{0}^a\! \left.\pder{\psi}{z} \right|_{z=0}^2 \!\!\! r \dint r \dint \theta + \frac{Um}{2},
\end{equation}
hence
\begin{equation}
E = \frac{\pi U^2 a^3}{2K} \!\!\sum_{n=0}^\infty \sum_{m=0}^\infty a_n a_m \!\!\int_0^1 \!\!\R_n(s) \R_m(s) s \dint s -\frac{a_0 \pi U^2 a^3}{8}  = \frac{\pi U^2 a^3}{8K} \!\!\left[ \sum_{n=0}^\infty \frac{a_n^2}{n+1} - a_0 K\right]\!.
\end{equation}
Note that as we are considering the limit of small $\mu$, the values of $a_n$ and $K$ are determined at $\mu = 0$.

For $\mu = 0$, $A_{mn} = \delta_{mn}/(4(n+1))$ and hence we may pre-multiply \cref{eq:eig1} by $\textbf{a}^T$ and divide through by $2K$ to show that
\begin{equation}
\frac{1}{8K}\left[ \sum_{n=0}^\infty \frac{a_n^2}{n+1}-a_0 K\right] = \frac{1}{2} a_m B_{mn} a_n.
\end{equation}
The right hand side of this result may alternatively be derived using the Plancherel Theorem for the Hankel transform
\begin{equation}
-\frac{1}{2}\int_0^\infty \left[\psi \pder{\psi}{z}\right]_{z = 0} \!\!r \dint r = \frac{U^2 a^3}{2}  \sin^2\theta \int_0^\infty \frac{A^2(\xi)}{\xi^2} \dint \xi = U^2 a^3 \sin^2\theta\,\, \frac{a_m B_{mn} a_n}{2},
\end{equation}
where the $\theta$ dependence may be integrated out to give the final factor of $\pi$. Further, we may show that
\begin{equation}
a_m B_{mn} a_n = -a_0,
\end{equation}
and hence
\begin{equation}
\label{eq:E_def}
E = -\frac{a_0 \pi U^2 a^3}{2},
\end{equation}
which is positive as $-a_0/2 \approx 3.10312$.


\section{The wave energy flux}
\label{sec:E_evol}

Here we derive the flux of energy from the vortex to the wavefield in the case of $\mu < 0$. Begin by considering the total energy contained within the region
\begin{equation}
\mathcal{V} = \{\textbf{x}: (x,y) \in \mathcal{A},\, z > 0\},
\end{equation}
where $\mathcal{A}$ is a finite, bounded region in 2D space with boundary $\partial\mathcal{A}$ and outward normal $\uv{n} \in \mathrm{span}\{\uv{x},\uv{y}\}$. The total energy within $\mathcal{V}$ is given by
\begin{equation}
E = \frac{1}{2}\int_\mathcal{V} |\nabla\psi|^2 \dint V,
\end{equation}
hence the rate of change of $E$ with time is given by
\begin{equation}
\label{eq:E_evol_1}
\dder{E}{t} = \int_\mathcal{V} \nabla \psi \cdot \nabla\psi_t \dint V = \int_\mathcal{V} \nabla\cdot\left[\psi\nabla\psi_t\right] \dint V = -\int_\mathcal{A} \left.\psi\pder{^2\psi}{z\partial t}\right|_{z=0} \!\!\!\!\dint A + \int_0^\infty\!\!\!\!\oint_{\partial\mathcal{A}} \psi \nabla\psi_t\cdot \uv{n} \dint l \dint z.
\end{equation}
We aim to convert all right hand side (RHS) integrals to boundary contributions describing the flux of energy through the boundary $\partial\mathcal{A}\times [0,\infty)$ so we examine the first RHS term of \cref{eq:E_evol_1} further.

On $z = 0$, from \cref{eq:SQG_BC} we have
\begin{equation}
-\pder{^2\psi}{z\partial t} = \nabla\cdot\left[\left(-U\uv{x}+\textbf{u}\right)\pder{\psi}{z} + \lambda\uv{x}\,\psi\right],
\end{equation}
for $\textbf{u} = (-\psi_y,\psi_x,0)$ and hence
\begin{equation}
-\psi\pder{^2\psi}{z\partial t} = \nabla\cdot\left[\left(-U\uv{x}+\textbf{u}\right)\psi\pder{\psi}{z} + \frac{\lambda}{2}\uv{x}\,\psi^2\right] + U\pder{\psi}{z}\pder{\psi}{x},
\end{equation}
on $z = 0$ so
\begin{equation}
\label{eq:E_evol_2}
-\int_\mathcal{A} \left.\psi\pder{^2\psi}{z\partial t}\right|_{z=0} \!\!\!\!\dint A = \oint_{\partial\mathcal{A}} \psi\left[\left(-U\uv{x}+\textbf{u}\right)\pder{\psi}{z} + \frac{\lambda}{2}\uv{x}\,\psi\right]_{z = 0}\!\!\!\!\!\!\!\cdot\uv{n}\dint l + \int_\mathcal{A} U\left.\pder{\psi}{z}\pder{\psi}{x}\right|_{z = 0}\!\!\!\! \dint A.
\end{equation}
The first term in \cref{eq:E_evol_2} is a boundary term of the required form, however, the second requires further work. 
The second RHS term of \cref{eq:E_evol_2} may be written as
\begin{equation}
\int_\mathcal{A} U\left.\pder{\psi}{z}\pder{\psi}{x}\right|_{z = 0}\!\!\!\! \dint A = \int_\mathcal{V} -\pder{}{z}\left[ \pder{\psi}{z}\pder{\psi}{x} \right] \dint V = -\int_\mathcal{V} \pder{\psi}{x}\pder{^2\psi}{z^2} + \frac{1}{2}\pder{}{x}\left[\pder{\psi}{z}\right]^2 \dint V,
\end{equation}
and substituting for $\pderline{^2\psi}{z^2}$ using \cref{eq:lap} allows us to convert all remaining terms to boundary contributions as
\begin{equation}
\int_\mathcal{A} U\left.\pder{\psi}{z}\pder{\psi}{x}\right|_{z = 0}\!\!\!\! \dint A 
= U \int_0^\infty\!\!\!\!\oint_{\partial\mathcal{A}} \left[- \frac{1}{2}|\nabla\psi|^2\uv{x} + \pder{\psi}{x}\nabla\psi\right]\cdot\uv{n} \dint l \dint z,
\end{equation}
where $\nabla_H = (\partial_x,\partial_y,0)$.

Our final energy decay is given by combining our results to get
\begin{multline}
\label{eq:E_evol_3}
\dder{E}{t} = \oint_{\partial\mathcal{A}} \psi\left[\left(-U\uv{x}+\textbf{u}\right)\pder{\psi}{z} + \frac{\lambda}{2}\uv{x}\,\psi\right]_{z = 0}\!\!\!\!\!\!\!\cdot\uv{n}\dint l \,\,+ \\ U \int_0^\infty\!\!\!\!\oint_{\partial\mathcal{A}} \left[- \frac{1}{2}|\nabla\psi|^2\uv{x} + \pder{\psi}{x}\nabla\psi\right]\cdot\uv{n} \dint l \dint z + \int_0^\infty\!\!\!\!\oint_{\partial\mathcal{A}} \psi \nabla\psi_t\cdot \uv{n} \dint l \dint z.
\end{multline}
To proceed, we take $\mathcal{V}$ to be a semi-infinite cylinder of radius $R \gg 1$ so $\partial\mathcal{A}$ is a circle of radius $R$ in the $(x,y)$ plane. Therefore, $\uv{n}\cdot\uv{x} = \cos\theta$ and $\uv{n}\cdot\nabla = \partial_r$ and hence
\begin{equation}
\label{eq:E_flux_0}
\dder{E}{t} = \int_0^{2\pi} \psi\left[-U\pder{\psi}{z}+ \frac{\lambda}{2}\psi\right]_{z = 0}\!\!\!\!\!\!\!R\cos\theta\dint \theta \,\,+ U \int_0^\infty\!\!\!\!\int_0^{2\pi} \left[- \frac{1}{2}|\nabla\psi|^2\cos\theta + \pder{\psi}{x}\pder{\psi}{r}\right]R \dint \theta \dint z, 
\end{equation}
where we have neglected terms cubic in $\psi$ as $\psi$ is small at large radius, $r = R$. Additionally, we have neglected the final integral of \cref{eq:E_evol_3} as the time derivative is small, $\partial_t = O(\mu^4)$, as shown above. Finally, \cref{eq:E_flux_0} may be written in cylindrical coordinates as
\begin{equation}
\label{eq:E_flux}
\dder{E}{t} = 
U \int_0^\infty\!\!\!\!\int_0^{2\pi} \left[ \frac{1}{2}\left( \left[\pder{\psi}{r}\right]^2-\left[\frac{1}{r}\pder{\psi}{\theta}\right]^2  \right)\cos\theta -\frac{1}{r}\pder{\psi}{\theta} \pder{\psi}{r} \sin\theta\right]_{r = R} \!\!\!\! R \dint \theta \dint z,
\end{equation}
where the first integral and the $z$ derivatives in \cref{eq:E_flux_0} are small, $O(\mu)$, and have been neglected.

\bibliographystyle{jfm}
\bibliography{bibliography}

\end{document}